\documentclass{article}
\usepackage[a4paper,margin={2.5cm}]{geometry}
\usepackage{graphicx}
\usepackage{amsmath}
\usepackage{amsfonts}
\usepackage{amssymb}
\usepackage{mathtools}
\usepackage{enumitem}
\usepackage{subcaption}
\usepackage{authblk}

\usepackage{ifthen}
\usepackage{colortbl}
\usepackage{array}
\usepackage{color}
\usepackage{soul}

\newboolean{IsWide}
\setboolean{IsWide}{true}

\newcommand{\FlaTwoByTwo}[4]{
\left(
\begin{array}{c I c}
#1 & #2 \\ \whline
#3 & #4
\end{array}
\right)
}

\newcommand{\FlaTwoByTwoSingleLine}[4]{
\left(
\begin{array}{c | c}
#1 & #2 \\ \hline
#3 & #4
\end{array}
\right)
}

\newcommand{\FlaTwoByOne}[2]{
\left(
\begin{array}{c}
#1 \\ \whline
#2
\end{array}
\right)
}

\newcommand{\FlaTwoByOneSingleLine}[2]{
\left(
\begin{array}{c}
#1 \\ \hline
#2
\end{array}
\right)
}

\newcommand{\FlaOneByTwo}[2]{
\left(
\begin{array}{c I c}
#1 & #2
\end{array}
\right)
}

\newcommand{\FlaOneByTwoSingleLine}[2]{
\left(
\begin{array}{c | c}
#1 & #2
\end{array}
\right)
}

\newcommand{\FlaThreeByThreeTL}[9]{
\left(
\begin{array}{c | c I c}
#1 & #2 & #3 \\ \hline
#4 & #5 & #6 \\ \whline
#7 & #8 & #9
\end{array}
\right)
}

\newcommand{\FlaThreeByThreeBR}[9]{
\left(
\begin{array}{c I c | c}
#1 & #2 & #3 \\ \whline
#4 & #5 & #6 \\ \hline
#7 & #8 & #9
\end{array}
\right)
}

\newcommand{\FlaOneByThreeR}[3]{
\left(
\begin{array}{c I c | c}
#1 & #2 & #3
\end{array}
\right)
}

\newcommand{\FlaOneByThreeL}[3]{
\left(
\begin{array}{c | c I c}
#1 & #2 & #3
\end{array}
\right)
}

\newcommand{\FlaThreeByOneT}[3]{
\left(
\begin{array}{c}
#1 \\ \hline
#2 \\ \whline
#3
\end{array}
\right)
}

\newcommand{\FlaThreeByOneB}[3]{
\left(
\begin{array}{c}
#1 \\ \whline
#2 \\ \hline
#3
\end{array}
\right)
}

\newcommand{\FlaPartition}[2]{
\ifthenelse{\boolean{IsWide}}{{\bf partition } \hspace{-1em} #1 \hspace{-1em} #2}
{{\bf partition } \+ \\ #1 \+ \\ #2 \- \-}
}

\newcommand{\FlaRepartition}[2]{
\ifthenelse{\boolean{IsWide}}{{\bf repartition } \hspace{-1em} #1 \hspace{-1em} #2}
{{\bf repartition } \+ \\ #1 \+ \\ #2 \- \-}
}

\newcommand{\FlaStartCompute}{%
\setlength{\unitlength}{1.6in}%
\begin{picture}(3,0.01)
\put(0,0){\line(1,0){3}}
\put(0,0.01){\line(1,0){3}}
\end{picture}%
}

\newcommand{\FlaEndCompute}{%
\noindent%
\setlength{\unitlength}{1.6in}%
\begin{picture}(3,0.01)
\put(0,0){\line(1,0){3}}
\put(0,0.01){\line(1,0){3}}
\end{picture}%
}

\newcommand{\operation}{ [ D, E, F, \ldots ] \becomes {\rm op}( A, B, C, D, \ldots ) }
\newcommand{\routinename}{ [ D, E, F, \ldots ] \becomes {\rm op}( A, B, C, D, \ldots ) }
\newcommand{\routinecost}{ X }
\newcommand{\precondition}{ Q }
\newcommand{\postcondition}{ R }
\newcommand{\invariant}{ P }
\newcommand{\costinv}{ \  }

\newcommand{\guard}{ R }
\newcommand{\partitionings}{
\begin{minipage}{3in}
$ S_I $
\end{minipage}
}
\newcommand{\initialize}{}

\newcommand{\partitionsizes}{ \hspace{ 3.25in} }

\newcommand{\blocksize}{1}

\newcommand{\repartitionings}{
\begin{minipage}[t]{3in}
\ \\
\ \\
\ \\
\end{minipage}
}

\newcommand{\repartitionsizes}{ \hspace{ 3.25in} }
\newcommand{\moveboundaries}{
\begin{minipage}[t]{3in}
\ \\
\ \\
\ \\
\end{minipage}
}
\newcommand{\beforeupdate}{
$ \QBefore $
}
\newcommand{\afterupdate}{
$ \QAfter $
}
\newcommand{\update}{%
\begin{minipage}[t]{4in}
$ S_U $
\end{minipage}
}

\newcommand{\resetsteps}{
\renewcommand{\blocksize}{1}
\renewcommand{\operation}{ [ D, E, F, \ldots ] \becomes {\rm op}( A, B, C, D, \ldots ) }
\renewcommand{\routinename}{ [ D, E, F, \ldots ] \becomes {\rm op}( A, B, C, D, \ldots ) }
\renewcommand{\routinecost}{ 0 }
\renewcommand{\precondition}{ \PPre }
\renewcommand{\postcondition}{ \PPost }
\renewcommand{\invariant}{ \PInv }
\renewcommand{\costinv}{ \  }
\renewcommand{\guard}{ G }
\renewcommand{\partitionings}{ %
\begin{minipage}[t]{3in}
\ \\
\end{minipage}
}
\renewcommand{\partitionsizes}{ \hspace{ 3.25in} }
\renewcommand{\repartitionings}{%
\begin{minipage}[t]{3in}
\ \\
\end{minipage}
}
\renewcommand{\repartitionsizes}{ \hspace{ 3.25in} }
\renewcommand{\moveboundaries}{%
\begin{minipage}[t]{3in}
\ \\
\end{minipage}
}
\renewcommand{\beforeupdate}{
\QBefore
}
\renewcommand{\afterupdate}{
\QAfter
}
\renewcommand{\update}{
$ S_U $
}
}

\newcommand{\WSguard}{
$ \guard $
}

\newcommand{\WSpartition}{%
\begin{minipage}[t]{3.0in}%
\begin{tabbing}
ind \= ind \= \kill
{\bf \color{blue} Partition}
\partitionings \+ \\
{\bf \color{blue} where } \hspace*{-2ex} \partitionsizes
\end{tabbing}
\end{minipage}
}

\newcommand{\WSinitialize}{
\vspace*{0.05cm}
\initialize
}

\newcommand{\WSrepartition}{
\begin{minipage}[t]{3in}
\ifthenelse{ \equal{\blocksize}{1} }{}
{%
\ifthenelse{ \equal{\blocksize}{blank} }{~}
{{\bf \color{blue} Determine block size} $ \blocksize $} \\
}
{\bf \color{blue} Repartition}
\vspace*{-0.2cm}
\begin{tabbing}
in \= in \= \+ \kill
\repartitionings \+ \\
{\bf \color{blue} where } \hspace*{-2ex} \repartitionsizes
\end{tabbing}
\end{minipage}
\vspace*{-0.3cm}
}

\newcommand{\WSrepartitionNarrow}{
\begin{minipage}[t]{2.05in}
\ifthenelse{ \equal{\blocksize}{1} }{\phantom{Determine}~}
{%
\ifthenelse{ \equal{\blocksize}{blank} }{}
{{\bf \color{blue} Determine block size} $ \blocksize $} \\
}
{\bf \color{blue} Repartition}
\begin{tabbing}
i \= i \= \+ \kill
\repartitionings 
\+ \\
{\bf \color{blue} where }
\begin{minipage}[t]{1.5in}
\repartitionsizes
\end{minipage}
\end{tabbing}
\end{minipage}
}

\newcommand{\WSmoveboundary}{%
\begin{minipage}[t]{4in}%
\vspace*{-0.2cm}
{\bf \color{blue} Continue with}
\vspace*{-0.2cm}
\begin{tabbing}
ind \= \+ \kill
\moveboundaries
\end{tabbing}
\end{minipage}
}

\newcommand{\WSupdate}{
\update
\vspace*{-0.3cm}
}

\newcommand{\FlaAlgorithmWithInit}{
\begin{center}
\begin{tabular}{| p{0.98\textwidth} |} \hline
{\bf \color{blue} Algorithm:} $\routinename$
\\ \whline
\WSinitialize \\
{\WSpartition} \\[0.3in]
\vspace*{-0.2cm}
{\bf \color{blue} while} \WSguard { \bf \color{blue} do} \\
\ \hspace{0.15in} \WSrepartition \\
\color{red} {\hspace{0.0in} \FlaStartCompute} \\
{\hspace{0.0in} \WSupdate} \\
\color{red} {\hspace{0.0in} \FlaEndCompute} \\
{\ \hspace{0.15in} \WSmoveboundary} \\
{{\bf \color{blue} endwhile}} \\
\vspace*{-0.3cm}
{{\bf \color{blue} end}} \\ \hline
\end{tabular}
\end{center}
}

\newcommand{\FlaAlgorithmWithoutLoop}{
\begin{center}
\begin{tabular}{| p{0.98\textwidth} |} \hline
{\bf \color{blue} Algorithm:} $\routinename$
\\ \whline
{\vspace*{0.05cm}
\hspace{0.0in} \WSupdate} \\
{{\bf \color{blue} end}} \\ \hline
\end{tabular}
\end{center}
}

\newcommand{ \PPre }{ P_{\it pre} }
\newcommand{ \PPost }{ P_{\it post} }
\newcommand{ \PInv }{ P_{\it inv} }

\newcommand{ \QBefore }{ P_{\it before} }
\newcommand{ \QAfter }{ P_{\it after} }

\newcommand{\Answer}[1]{%
\ifthenelse{\boolean{ShowAnswers}}{%
{\color{blue}%
\vspace{0.05in}%
\noindent%
{\bf Answer:} #1}}
{}%
}

\newcommand{\PartAnswer}[1]{%
\ifthenelse{\boolean{ShowAnswers}}{%
{\color{blue} #1}}%
{}
}

\newcommand{\QuestionAnswer}[2]{%
\ifthenelse{\boolean{ShowAnswers}}{%
{\color{blue} #2}}%
{\color{black} #1}
}

\newcommand{\ShowAnswer}[1]{
\ifthenelse{\boolean{ShowAnswers}}{
{\color{blue} #1}}
{\phantom{#1}}
}

{
\vspace{0.1in}
\end{minipage}
\\ \whline
\end{tabular}
}

\newenvironment{unboxedexercise}{
\addtocounter{homeworkcounter}{1}%
\noindent%
\begin{exercise}
\rm 
}
{
\end{exercise}%
}

\newcolumntype{I}{!{\vrule width 1.5pt}}
\newlength\savedwidth
\newcommand\whline{\noalign{\global\savedwidth\arrayrulewidth
                            \global\arrayrulewidth 1.5pt}%
           \hline
           \noalign{\global\arrayrulewidth\savedwidth}}

\newcommand{\NoShow}[1]{}

\newcommand{\becomes}{:=}

\definecolor{color0}{RGB}{117,195,255}
\definecolor{color1}{RGB}{0,0,255}
\definecolor{color2}{RGB}{255,255,255}
\definecolor{color3}{RGB}{255,0,0}
\definecolor{color4}{RGB}{255,0,174}
\definecolor{color5}{RGB}{179,0,0}
\definecolor{color6}{RGB}{0,255,0}
\definecolor{color7}{RGB}{255,255,0}
\definecolor{color8}{RGB}{235,0,0}
\definecolor{color9}{RGB}{0,162,0}
\definecolor{color0}{RGB}{255,0,255}
\definecolor{color11}{RGB}{100,100,177}
\definecolor{color12}{RGB}{172,174,41}
\definecolor{color13}{RGB}{255,144,26}
\definecolor{color14}{RGB}{2,255,177}
\definecolor{color15}{RGB}{192,224,0}
\definecolor{color16}{RGB}{66,66,66}
\definecolor{color17}{RGB}{255,0,96}
\definecolor{color18}{RGB}{169,169,169}
\definecolor{color19}{RGB}{169,0,0}
\definecolor{color20}{RGB}{0,109,255}
\definecolor{color21}{RGB}{200,61,68}
\definecolor{color22}{RGB}{200,66,0}
\definecolor{color23}{RGB}{0,41,0}
\definecolor{color24}{RGB}{139,121,177}
\definecolor{color25}{RGB}{116,116,116}
\definecolor{color26}{RGB}{200,50,89}
\definecolor{color27}{RGB}{255,171,98}
\definecolor{color28}{RGB}{0,68,189}
\definecolor{color29}{RGB}{52,43,0}
\definecolor{color30}{RGB}{255,46,0}
\definecolor{color31}{RGB}{100,216,32}

\newtheorem{remark}{Remark}

\newcommand{\defeq}{\vcentcolon=}
\newcommand{\si}{(i)}    
\newcommand{\simo}{(i-1)} 

\newlength{\topFigureVerticalSpace}
\newlength{\bottomFigureVerticalSpace}
\newcommand{\tfvspace}{\vspace*{\topFigureVerticalSpace}}
\newcommand{\bfvspace}{\vspace*{-\bottomFigureVerticalSpace}}

\newcommand{\mtx}[1]{#1}

\begin{document}

\title{Solving Large Rank-Deficient Linear Least-Squares Problems
       on Shared-Memory CPU Architectures and GPU Architectures}


\author[1]{M\'onica Chillar\'on}
\affil[1]{
  Dpto.~de Sistemas Informáticos y Computaci\'on,
  Universitat Polit\`ecnica de Val\`encia, 
  46022--Val\`encia, Spain.
  email: \texttt{mnichipr@inf.upv.es}
}

\author[2]{Gregorio Quintana-Ort\'{i}}
\affil[2]{%
  Dpto.~de Ingenier\'{\i}a y Ciencia de Computadores,
  Universitat Jaume I de Castell\'on,
  12071--Castell\'on, Spain.
  email: \texttt{gquintan@uji.es}
}  

\author[3]{Vicente Vidal}
\affil[3]{%
  Dpto.~de Sistemas Informáticos y Computaci\'on,
  Universitat Polit\`ecnica de Val\`encia, 
  46022--Val\`encia, Spain.
  email: \texttt{vvidal@dsic.upv.es}
}  

\author[4]{Per-Gunnar Martinsson}
\affil[4]{%
  Dept.~of Mathematics, 2515 Speedway,
  University of Texas at Austin,
  78712, Austin, Texas, USA.
  email: \texttt{pgm@oden.utexas.edu}
}

\setcounter{Maxaffil}{0}
\renewcommand\Affilfont{\itshape\small}

\maketitle

\begin{abstract}

Solving very large linear systems of equations
is a key computational task in science and technology.
In many cases, the coefficient matrix of the linear system is rank-deficient, 
leading to systems that may be underdetermined, inconsistent, or both. 
In such cases, 
one generally seeks to compute the least squares solution 
that minimizes the residual of the problem,
which can be further defined as the solution with smallest norm
in cases where the coefficient matrix has a nontrivial nullspace.
This work presents several new techniques for solving least squares
problems involving coefficient matrices that are
so large that they do not fit in main memory.
The implementations include both CPU and GPU variants.
All techniques rely on complete orthogonal decompositions
that guarantee that both conditions of a least squares solution
are met, regardless of the rank properties of the matrix.
Specifically, they rely on the recently proposed ``randUTV'' 
algorithm that is particularly effective in strongly 
communication-constrained environments.
A detailed precision and performance study reveals that
the new methods, that operate on data stored on disk,
are competitive with state-of-the-art methods that
store all data in main memory.

\end{abstract}


\section{Introduction}


Let $A$ be a matrix of dimension $m \times n$ and let $b$ be a vector of dimension $m$.
Let $r$ denote the rank of the matrix to some given computational precision $\tau$.
In solving a linear system  $Ax = b$, two complications may arise.
If the rank $r$ is smaller than $m$, it may be the case that the data vector $b$
does not belong to the range of $A$, in which case the system is \textit{inconsistent}
since it does not admit an exact solution.
If the rank $r$ is smaller than $n$, then the matrix has a non-trivial nullspace,
and the solution is not unique.
When either or both complications are present, 
we generally look for the \textit{least squares solution} $x$, 
which is defined by two conditions: 
(1) It minimizes the residual $\|Ax - b\|$ over $x \in \mathbb{R}^{n}$, and 
(2) among the minimizers, the solution $x$ is the one with smallest magnitude.
Different norms may be considered, but the Euclidean norm is the
most common, and is the only one considered in this work.

In cases where the coefficient matrix has no nullspace
(i.e.~the matrix has full column rank),
least squares problems may be solved very efficiently
by using a plain (unpivoted) QR factorization~\cite{golub}
based on Householder transformations,
which provide a good numerical stability.

However, for more general problems, state-of-the-art software is based
on the singular value decomposition (SVD) \cite{golub1965calculating}, 
which ensures that both conditions of a least square solution are met, 
regardless of the aspect ratio and rank of the matrix. 
For computational efficiency, 
high-performance software often relies on 
\textit{complete orthogonal decompositions} (CODs)~\cite{hough1997complete},
which generalize the SVD to build a decomposition of the form
$\mtx{A} = \mtx{U}\mtx{T}\mtx{V}^{*}$, 
where $\mtx{U}$ and $\mtx{V}$ are unitary
and $\mtx{T}$ is triangular. 
Such decompositions can be faster than the SVD, 
and are sufficient for computing the least square solution
in a numerically stable way. 
One of the most practical methods for computing a COD is based
on the column-pivoted QR factorization (CPQR),
which requires a much lower computational cost than the SVD
and usually reveals rank reasonably well.
%
The original method was developed by Golub and Kahan, and
was later optimized and refined for 
unicore architectures and 
shared-memory multicore/multiprocessor 
architectures~\cite{Businger:1965:LLS,quintana1998blas,Drmac2008}.
Nevertheless, 
least squares solvers based on both SVD and CPQR
are much slower than least-squares methods for full rank matrices
based on the unpivoted QR factorization on current architectures
because of the large ratio of matrix-vector operations of 
both the SVD and CPQR.

When working with matrices so large that they must be stored 
in an external device (also called stored Out-Of-Core or OOC),
both least-squares solvers that employ the SVD and 
least-squares solvers that employ CPQR
are strongly limited in performances
because of the many slow matrix-vector operations required
and the large amount of data being transferred.
Some efforts for computing factorizations of dense OOC matrices include
POOCLAPACK~\cite{gunter2005parallel,reiley1999pooclapack,gunter2001parallel},
the SOLAR library~\cite{toledo1996design},
a runtime to execute algorithms-by-blocks~\cite{quintana2012oocruntime},
as well as an OOC extension~\cite{d2000design} to ScaLAPACK 
not included in this library.
These efforts usually are limited to Cholesky, unpivoted QR, and
LU decompositions.
A truncated SVD algorithm that is effective for out-of-core matrices
was introduced by Halko \textit{et al.}~\cite{halko2011algorithm}.
This technique, however, only yields a \textit{partial} factorization, and
it requires upper bound for the target rank $k$ known in advance.
Recent works have been focused on computing the SVD of dense OOC matrices.
Demchik \textit{et al.}~\cite{demchik:2019:svd}
employed randomization to compute the SVD,
but no performance results for dense matrices were reported.
Kabir \textit{et al.}~\cite{kabir:2017:svd}
used a two-stage approach for computing the SVD
that reduces the dense matrix to band form in a first stage,
which allowed to factorize
matrices of up to dimension $100k \times 100k$.

In cases where the coefficient matrix is sparse but of high dimension, 
iterative techniques can be employed,
such as LSQR~\cite{paige1982lsqr}, 
LSMR~\cite{fong2011lsmr}, 
LSMB~\cite{hallman2018lsmb},
and other Krylov methods~\cite{saad2003iterative}. 
These methods have the advantage of iterating with the coefficient matrix without modifying it. Therefore, the required memory resources are lower, although the convergence rate can also be low. To accelerate convergence, they can be combined with preconditioners or regularization techniques~\cite{beck2009fast, fornasier2016conjugate, scott2017using, scott2017solving}. 
Recently, a method 
for computing preconditioners 
based on Krylov's methods~\cite{cerdan2020preconditioners} was presented, 
and in most of the tests 
applying preconditioners accelerates the convergence 
of the iterative Krylov subspace method LSMR 
when having a rank-deficient problem.

Among the direct methods for sparse matrices, 
the QR approach is the most used, 
and there are several libraries that offer optimized methods, 
such as Intel's MKL Sparse QR~\cite{fedorov-2018}, 
SuiteSparse QR~\cite{foster2013algorithm, yeralan2017algorithm}, and 
the MUMPS QR solver~\cite{agullo2015task}. 
Another option for large sparse systems is the Multilevel Incomplete QR preconditioning technique, 
included in the MIQR package~\cite{li2006miqr}. 
However, all of them are focused on sparse matrices. 
If a part of the sparse matrix is dense, 
the factorization will fail if the main memory is not large enough. 
A recent work~\cite{scott2022computational} proposes 
a hybrid approach to solve the problem if the matrix has some dense rows. 
Scott and Tuma divide the matrix into two parts: 
the sparse part and the dense part. 
They perform an updating, partial stretching or regularization on the matrix. 
Then, they use either the QR factorization or a hybrid QR-iterative method 
to solve the problem. 
This method is more robust and efficient, and 
the sparse part of  the matrix can be rank deficient. 
Nevertheless, the main problem is that it is still proposed 
for mostly-sparse matrices, and 
memory resources may be insufficient when the problem is big enough or not sparse enough.

Some methods, such as the Pivoting Avoiding QR (PAQR)~\cite{paqr2023}, 
can work with dense matrices as long as they fit in main memory. 
PAQR is a variant of the standard QR factorization 
that eliminates the need for explicit pivoting, 
making it a more efficient option for matrices with significant linear dependence. 
The key to the PAQR algorithm is the dynamic detection and elimination of 
linearly dependent columns during the factorization process. 
Instead of pivoting to remove dependent columns, 
PAQR identifies these columns and removes them directly from the process.
This reduces the number of operations required and improves the performance 
of the algorithm, especially for matrices with a large number of linearly dependent columns. 
The results show that PAQR is significantly more efficient 
than QR for matrices with significant linear dependence.

However, in order to solve very large dense or not-so-sparse systems, 
other techniques must be employed. As nowadays disk space is much cheaper than main memory,
an interesting approach is 
using Out-Of-Core techniques~\cite{gunter2005parallel, quintana2012runtime}.

The key contribution of the present manuscript is 
a set of algorithms and implementations for solving least squares problems 
via complete orthogonal decompositions when
the coefficient matrix is too large to fit in main memory, and is instead stored
``Out-Of-Core'' (OOC) on cheaper disk space. 
Our work builds on prior least squares solvers for this environment 
that rely on the QR decomposition 
(making them suitable only for problems with full column rank) 
for both CPU and GPU architectures
(\cite{chillaron2020computed} and \cite{quintana2022high}, respectively).
The maximum matrix size assessed in those works 
was $266\,500 \times 262\,144$,
which requires about 520 GB of storage 
when working with double-precision arithmetic, 
making it infeasible for methods 
that work with data in main memory (``in-core'' computations) 
on most computers. 

To enable the solution of least squares problems 
involving coefficient matrices of arbitrary sizes and arbitrary ranks, 
the QR factorization is swapped out in favor of
the ``randUTV'' algorithm \cite{martinsson2017randutv},
which can be viewed as a blocked incremental version of 
the randomized SVD (RSVD)~\cite{2011_martinsson_randomsurvey,2007_martinsson_PNAS}. 
The randUTV factorization can be implemented with high performances
when data is stored Out-Of-Core on CPU architectures~\cite{2023_martinsson_randUTV_ooc}. 
In our work, 
this randUTV implementation has been accelerated with some improvements,
and then the technique is further extended to solve the linear least squares problem
by incorporating the following three tasks:
The first task is a simple rank estimation step.
The second task is a computation that modifies 
the basis for the row space (encoded in the columns of $V$) 
so that it provides a clear split 
between the nullspace of the matrix and its orthogonal complement. 
This second task is necessary to ensure that a minimal norm solution is constructed.
The third task is the final computations to obtain the solution $x$.

\vspace{2mm}

\textit{\textbf{Remark:}
The manuscript also describes a version of the code that skips the second step,
and relies instead on a direct truncation of the factorization resulting
from the randUTV algorithm. This version computes a solution that minimizes the 
residual (up to floating-point precision), but is not guaranteed to result in a
solution with minimal norm. However, since randUTV by itself is highly accurate
in revealing the numerical rank, the loss of optimality is very minor, as demonstrated
in several numerical examples.
}

\vspace{2mm}


The main contributions of the manuscript are enumerated next:

\begin{itemize}

\item
We present a new set of implementations 
for solving linear least-squares problems
so large that they do not fit in main memory.

\item
Our new methods can solve linear systems
with both full-rank and rank-deficient coefficient matrices,
for both consistent and inconsistent problems.

\item
Our new methods work for any matrices that fit on the disk drive,
regardless of whether they fit in main memory.

\item
Our work accelerates the speed of previous Out-Of-Core implementations
of the randUTV factorization for CPU architectures
by introducing several performance improvements.
Moreover, we developed a new family of implementations for GPU architectures.

\item
A precision study of our new implementations has been performed,
which includes both our own matrices and
several large matrices from public repositories.
The precision of the new codes is competitive
with those of methods working in main memory.

\item
A thorough performance study has been performed.
The performances of the new codes are competitive with 
those of state-of-the-art high-performance codes with similar functionality 
that work only on data stored in main memory 
(thus limited in size).

\item
The performances of our new codes are slower
than those of specific methods for full-rank matrices,
but only slightly so.

\item
Our new methods can be employed both in CPU architectures
and GPU architectures,
delivering good performance and precision in both cases.

\end{itemize}

The document is organized as follows: 
Section~\ref{sec:methods} describes the traditional algorithms 
for solving the linear least squares problem.
Section~\ref{sec:randutv} introduces our new approach
for solving the linear least squares problem
by using the randUTV factorization.
Section~\ref{sec:numerical_analysis} contains 
both a performance assessment and a precision study 
comparing our implementations 
with respect to high-performance implementations 
employed when the data fits in main memory.
Finally, Section~\ref{sec:conclusions} enumerates 
the conclusions of our work.

\section{Methods for Solving Linear Least-Squares Problems
         with Dense Coefficient Matrices}
\label{sec:methods}

In this section, we briefly review the classical least squares problem and
describe the principal algorithms for solving it.
Throughout the section, $A$ denotes a matrix of size $m\times n$, $b$ denotes
an $m\times 1$ vector, and we seek to solve the linear system
$$
Ax = b
$$
in a least squares sense. To be precise, we seek the $x$ vector 
that satisfies the following two conditions:
\begin{itemize}
\item[(1)] The vector $x$ minimizes the residual. In other words
$$
\|Ax - b\| = \inf\{\|Ay - b\|\,\colon\,y \in \mathbb{R}^{n}\}.
$$
\item[(2)] If the problem in (1) has more than one solution, then $x$ is the
solution with minimal norm.
\end{itemize}
Throughout the manuscript, $\|\cdot\|$ denotes Euclidean norm, unless otherwise noted.

\subsection{Methods based on the unpivoted QR decomposition}

When $A$ has full column rank (so that $\mbox{rank}(A)=n$), condition (2) is not
operable, and the least squares solution can be computed via an unpivoted QR
decomposition. 
In this case, 
the first step is to build the factorization $A = QR$,
where $Q \in \mathbb{R}^{m\times n}$ has orthonormal columns and 
$R \in \mathbb{R}^{n\times n}$ is upper triangular.
The least squares solution is then $x = R^{-1}\bigl(Q^{*}b\bigr)$.
Note that since $R$ is triangular,
there is no need to compute the explicit inverse of $R$.
Also note that the matrix $Q$ does not need to be explicitly built,
and can be applied to $b$ on-the-fly when factorizing the matrix $A$.

\subsection{Methods based on the singular value decomposition}
\label{subsec:SVD}

In the case where the coefficient matrix does not have full column rank,
the standard tool for solving the least squares problem is based on truncating
the singular value decomposition (SVD). 
In this approach, one first computes a factorization 
$A = U\Sigma V^{*}$, 
where $U$ and $V$ are unitary and $\Sigma$ is diagonal. 
The next step is to determine the numerical rank $r$ of the matrix 
to some given tolerance $\tau$, and partition the factors so that
$$
A = \bigl[U_{1}\quad U_{2}\bigr]
\left[\begin{array}{rr} \Sigma_{11} & 0 \\ 0 & 0 \end{array}\right]
\left[\begin{array}{r} V_{1}^{*} \\ V_{2}^{*} \end{array}\right],
$$
where $\Sigma_{11}$ is of size $r\times r$. 
The least squares solution is finally
$x = V_{1}\Sigma_{11}^{-1}\bigl(U_{1}^{*}b\bigr)$.

\subsection{Methods based on complete orthogonal decompositions}
\label{subsec:COD}

Techniques based on the SVD form the gold standard 
for solving least squares problems. 
Due to the high cost of computing a full SVD, 
a simplified procedure based on complete orthogonal decompositions (CODs) 
is sometimes used. 
The idea is to form a factorization $A = UTV^{*}$,
where $U$ and $V$ are unitary and $T$ is triangular. 
Concretely, 
the SVD is a special case of a COD, and the reason to introduce
additional flexibility in the middle factor is that it enables simpler and faster
algorithms. 

The standard approach for computing a COD is 
via two orthogonal factorizations.
The first one computes $AV = UT$,
where $V$ is a permutation matrix,
$U$ is unitary, 
and $T$ is upper triangular.
The rank $r$ is detected as the number of non-zero rows in $T$, so that the
factorization takes the form
$$
AV =
\bigl[U_{1}\quad U_{2}\bigr]
\left[\begin{array}{rr} T_{11} & T_{12} \\ 0 & 0 \end{array}\right] = 
U_{1}\bigl[T_{11},\,T_{12}\bigr],
$$
where $T_{11}$ is of size $r \times r$. 

Then, a second orthogonal factorization is used to build an orthonormal basis 
for the row space of $A$ through a factorization
$\bigl[ T_{11}, T_{12} \bigr] = S_{11}^{*}W^{*}$,
where $W \in \mathbb{R}^{n\times r}$ has orthonormal columns and
$S_{11} \in \mathbb{R}^{r\times r}$ is upper triangular. In other words,
we build the factorization
$
AV = U_{1}S_{11}^{*}W^{*}.
$
The least square solution is then
$
x = V\,W\bigl(S_{11}^{*}\bigr)^{-1}(U_{1}^{*}b).
$
Note that in practice the matrix $A$ is typically rank deficient only
to some preset tolerance $\tau$.

\section{Solving the linear least-squares problem
         with the randUTV factorization}
\label{sec:randutv}

The methods we propose follow the traditional template of methods
based on complete orthogonal decompositions described 
in Subsection~\ref{subsec:COD}.
The key difference is that
the first step of our process uses the ``randUTV'' algorithm,
which given a matrix $A \in \mathbb{R}^{m\times n}$
computes a decomposition
\begin{equation}
\label{eq:UTVdef}
\begin{array}{ccccccccccc}
A & V &=& U & T & ,\\
m\times n & n\times n && m\times m & m\times n
\end{array}
\end{equation}
where $U$ and $V$ are unitary, and $T$ is upper triangular. The key difference to
the traditional techniques is that we allow $V$ to involve unitary transformations
beyond column permutations. This additional freedom has two benefits: 
The first one is that 
it allows the algorithm to be \textit{blocked}, 
which is essential for high-performance Out-Of-Core implementations 
like the one described here.
The second one is that
it allows for much stronger rank-revealing properties, in the sense that most
of the mass of $T$ gets moved to the diagonal, with off-diagonal entries being much
smaller in magnitude than in column-pivoted QR (CPQR).
This second benefit is important 
in the present context 
since it allows us to estimate the numerical rank with a higher accuracy.

In this section, we briefly review randUTV, 
and then describe the details of how we implement it in the present context.

\subsection{The randUTV algorithm}

The randUTV algorithm takes as its input 
a general matrix $\mtx{A} \in \mathbb{R}^{m\times n}$ and 
a block size $n_b$, 
and drives $A$ to upper triangular form 
via a sequence of $s= \lceil n/n_b \rceil$ unitary transformations. 
(We refer to the matrix $T$ as ``upper triangular'' 
even in the case $n < m$, where $T$ is technically ``trapezoidal''.) 
We start with $\mtx{T}^{(0)} \defeq \mtx{A}$. In the $i$-th iteration ($i=1,2,\ldots,s$), a matrix $\mtx{T}^{\si} \in \mathbb{R}^{m \times n}$ is formed by the computation
\begin{equation}
\mtx{T}^{\si} \defeq \left(\mtx{U}^{\si}\right)^* \mtx{T}^{\simo} \mtx{V}^{\si},
\end{equation}
where $\mtx{U}^{\si} \in \mathbb{R}^{m \times m}$ and $\mtx{V}^{\si} \in \mathbb{R}^{n \times n}$ are unitary 
matrices chosen to introduce zeros under the diagonal of the matrix, and to promote the rank-revealing quality
of the decomposition. 
Figure~\ref{fig:randutv-pattern} shows 
the general pattern of the transformations for a toy example.


\begingroup

\setlength{\tabcolsep}{1.8pt}
\renewcommand{\arraystretch}{0.92}

\newcommand{\newvs}{\vspace*{0.4cm}}
\newcommand{\newminusvs}{\vspace*{-0.1cm}}

\newcommand\sbullet[1][.5]{\mathbin{\vcenter{\hbox{\scalebox{#1}{$\bullet$}}}}}

\newcommand{\mybu}{$\bullet$}
\newcommand{\mylb}{$\sbullet[1.25]$}
\newcommand{\mysb}{$\sbullet[0.6]$}



\captionsetup[subfigure]{labelformat=empty}

\begin{figure}[ht!]
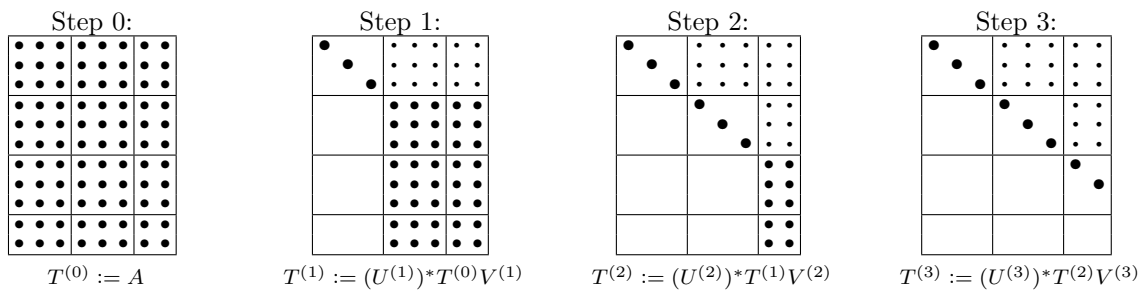

\begin{subfigure}[t]{0.24\linewidth}
\scriptsize
\centering
\begin{tabular}{|ccc|ccc|cc|}
  \multicolumn{8}{c}{\normalsize Step 0:} \\ \hline
  \mybu & \mybu & \mybu & \mybu & \mybu & \mybu & \mybu & \mybu \\ 
  \mybu & \mybu & \mybu & \mybu & \mybu & \mybu & \mybu & \mybu \\
  \mybu & \mybu & \mybu & \mybu & \mybu & \mybu & \mybu & \mybu \\ \hline
  \mybu & \mybu & \mybu & \mybu & \mybu & \mybu & \mybu & \mybu \\
  \mybu & \mybu & \mybu & \mybu & \mybu & \mybu & \mybu & \mybu \\
  \mybu & \mybu & \mybu & \mybu & \mybu & \mybu & \mybu & \mybu \\\hline
  \mybu & \mybu & \mybu & \mybu & \mybu & \mybu & \mybu & \mybu \\
  \mybu & \mybu & \mybu & \mybu & \mybu & \mybu & \mybu & \mybu \\
  \mybu & \mybu & \mybu & \mybu & \mybu & \mybu & \mybu & \mybu \\\hline
  \mybu & \mybu & \mybu & \mybu & \mybu & \mybu & \mybu & \mybu \\
  \mybu & \mybu & \mybu & \mybu & \mybu & \mybu & \mybu & \mybu \\ \hline
\end{tabular}
\newminusvs
\caption{\footnotesize $\mtx{T}^{(0)} \defeq \mtx{A}$}
\newvs
\end{subfigure}
\hspace*{0.00cm}
\begin{subfigure}[t]{0.24\linewidth}
\scriptsize
\centering
\begin{tabular}{|ccc|ccc|cc|}
  \multicolumn{8}{c}{\normalsize Step 1:} \\ \hline
  \mylb &       &       & \mysb & \mysb & \mysb & \mysb & \mysb \\ 
        & \mylb &       & \mysb & \mysb & \mysb & \mysb & \mysb \\
        &       & \mylb & \mysb & \mysb & \mysb & \mysb & \mysb \\ \hline
        &       &       & \mybu & \mybu & \mybu & \mybu & \mybu \\
        &       &       & \mybu & \mybu & \mybu & \mybu & \mybu \\
        &       &       & \mybu & \mybu & \mybu & \mybu & \mybu \\\hline
        &       &       & \mybu & \mybu & \mybu & \mybu & \mybu \\
        &       &       & \mybu & \mybu & \mybu & \mybu & \mybu \\
        &       &       & \mybu & \mybu & \mybu & \mybu & \mybu \\\hline
        &       &       & \mybu & \mybu & \mybu & \mybu & \mybu \\
        &       &       & \mybu & \mybu & \mybu & \mybu & \mybu \\ \hline
\end{tabular}
\newminusvs
\caption{\footnotesize $\mtx{T}^{(1)} \defeq (\mtx{U}^{(1)})^* \mtx{T}^{(0)} \mtx{V}^{(1)}$}
\newvs
\end{subfigure}
\hspace*{0.00cm}
\begin{subfigure}[t]{0.24\linewidth}
\scriptsize
\centering
\begin{tabular}{|ccc|ccc|cc|}
  \multicolumn{8}{c}{\normalsize Step 2:} \\ \hline
  \mylb &       &       & \mysb & \mysb & \mysb & \mysb & \mysb \\ 
        & \mylb &       & \mysb & \mysb & \mysb & \mysb & \mysb \\
        &       & \mylb & \mysb & \mysb & \mysb & \mysb & \mysb \\ \hline
        &       &       & \mylb &       &       & \mysb & \mysb \\
        &       &       &       & \mylb &       & \mysb & \mysb \\
        &       &       &       &       & \mylb & \mysb & \mysb \\\hline
        &       &       &       &       &       & \mybu & \mybu \\
        &       &       &       &       &       & \mybu & \mybu \\
        &       &       &       &       &       & \mybu & \mybu \\\hline
        &       &       &       &       &       & \mybu & \mybu \\
        &       &       &       &       &       & \mybu & \mybu \\ \hline
\end{tabular}
\newminusvs
\caption{\footnotesize $\mtx{T}^{(2)} \defeq (\mtx{U}^{(2)})^* \mtx{T}^{(1)} \mtx{V}^{(2)}$}
\newvs
\end{subfigure}
\hspace*{0.00cm}
\begin{subfigure}[t]{0.24\linewidth}
\scriptsize
\centering
\begin{tabular}{|ccc|ccc|cc|}
  \multicolumn{8}{c}{\normalsize Step 3:} \\ \hline
  \mylb &       &       & \mysb & \mysb & \mysb & \mysb & \mysb \\ 
        & \mylb &       & \mysb & \mysb & \mysb & \mysb & \mysb \\
        &       & \mylb & \mysb & \mysb & \mysb & \mysb & \mysb \\ \hline
        &       &       & \mylb &       &       & \mysb & \mysb \\
        &       &       &       & \mylb &       & \mysb & \mysb \\
        &       &       &       &       & \mylb & \mysb & \mysb \\\hline
        &       &       &       &       &       & \mylb &       \\
        &       &       &       &       &       &       & \mylb \\
        &       &       &       &       &       &       &       \\\hline
        &       &       &       &       &       &       &       \\
        &       &       &       &       &       &       &       \\ \hline
\end{tabular}
\newminusvs
\caption{\footnotesize $\mtx{T}^{(3)} \defeq (\mtx{U}^{(3)})^* \mtx{T}^{(2)} \mtx{V}^{(3)}$}
\newvs
\end{subfigure}
\vspace*{-0.4cm}
\caption{The sparsity patterns of the four matrices $\mtx{T}^{\si}$ that
appear in randUTV, shown for the particular case where $m=11, n=8$, and $n_b=3$.
The size of each element is a rough approximation of its absolute value.
}
\label{fig:randutv-pattern}
\end{figure}

\endgroup

The theoretically ideal UTV decomposition is one where the first $j$ columns of $U$/$V$ span the same space as the dominant
$j$ left/right singular vectors of $A$. 
The randUTV algorithm gets close to this optimum 
through a randomized subspace approximation 
that proceeds in blocks of $n_b$ vectors. 
To be precise, in the first step of randUTV, 
a matrix $G \in \mathbb{R}^{m\times n_b}$ is drawn from a Gaussian distribution, 
and a sample matrix $Y = A^{*}G$ is computed. 
The first $n_b$ columns of $V$ are then constructed 
to form an orthonormal (ON) basis 
for the space spanned by the  columns of $Y$, 
and the first $n_b$ columns of $U$ are constructed 
to form an ON basis for the space spanned 
by the columns of $AV(:,1:n_b)$. 
The block $T(1:n_b,1:n_b)$ is diagonal, 
and holds the $n_b$ singular values of $Y = A^{*}G$. 

To further strengthen the rank revealing properties of randUTV, 
a small number of power iterations can be
incorporated into the randomized sampling, 
so that we instead build 
the sample matrix $Y = (A^{*}A)^{q}A^{*}G$ for some small integer $q$. 
In practice, $q=1$ or $q=2$ is sufficient to yield very high accuracy
(the diagonals approximate the singular values with several digits of precision). 

Once the first step of randUTV has been completed, 
the first $n_b$ columns of the initial matrix $T^{(0)} = A$ 
have been driven to upper triangular form in the updated matrix $T^{(1)} = (U^{(1)})^{*}T^{(0)}V^{(1)}$.
The process is then repeated on the matrix formed by excluding the first $n_b$ columns and rows of $T^{(1)}$,
and continues through a sequence of unitary transforms until an upper triangular matrix has been obtained.

Figure~\ref{fig:alg_utv} shows the complete randUTV algorithm 
written with the FLAME methodology/notation.
In this algorithm,
the second output argument of the unpivoted QR factorization
is the upper triangular factor, whereas 
the third one is the unit lower trapezoidal matrix 
with the Householder vectors.

\begingroup

\setlength{\arraycolsep}{3pt}
\resetsteps 


\renewcommand{\routinename}{ 
  \left[ U,T,V \right] = \mbox{\sc randUTV}( A, q, n_b )
}


\renewcommand{\initialize}{
  $
    V \defeq \mbox{\sc eye}(n(A), n(A)), $\quad$
    U \defeq \mbox{\sc eye}(m(A), m(A))
  $
}


\renewcommand{\partitionings}{
  $
  A \rightarrow
  \FlaTwoByTwo{A_{TL}}{A_{TR}}
              {A_{BL}}{A_{BR}}
  $
  ,
  $
  V \rightarrow
  \FlaOneByTwo{V_{L}} {V_{R}}
  $
  ,
  $
  U \rightarrow
  \FlaOneByTwo{U_{L}}
              {U_{R}}
  $
}

\renewcommand{\partitionsizes}{
  $\:$
  $ A_{TL} $ is $ 0 \times 0 $,
  $ V_{L} $ has $ 0 $ columns,
  $ U_{L} $ has $ 0 $ columns
}


\renewcommand{\guard}{
  m( A_{TL} ) < m( A )
}


\renewcommand{\blocksize}{b=\min(n_b,n(A_{BR}))}

\renewcommand{\repartitionings}{
  $  
    \FlaTwoByTwo{A_{TL}}{A_{TR}}
                {A_{BL}}{A_{BR}}
    \rightarrow
    \FlaThreeByThreeBR{A_{00}}{A_{01}}{A_{02}}
                      {A_{10}}{A_{11}}{A_{12}}
                      {A_{20}}{A_{21}}{A_{22}}
  $
  , \\
  $
    \FlaOneByTwo{ V_L } { V_R }
    \rightarrow
    \FlaOneByThreeR{V_0} {V_1} {V_2}
  $
  , 
  $  
    \FlaOneByTwo{ U_L } { U_R }
    \rightarrow
    \FlaOneByThreeR{U_0} {U_1} {U_2}
  $
}

\renewcommand{\repartitionsizes}{
  $\:$
  $ A_{11} $ is $ n_b \times n_b $,
  $ V_1 $ has $ n_b $ rows,
  $ U_1 $ has $ n_b $ rows
}


\renewcommand{\moveboundaries}{
  $  
    \FlaTwoByTwo{A_{TL}}{A_{TR}}
                {A_{BL}}{A_{BR}}
    \leftarrow
    \FlaThreeByThreeTL{A_{00}}{A_{01}}{A_{02}}
                      {A_{10}}{A_{11}}{A_{12}}
                      {A_{20}}{A_{21}}{A_{22}}
  $
  , \\
  $
    \FlaOneByTwo{ V_L } { V_R }
    \leftarrow
    \FlaOneByThreeL{V_0} {V_1} {V_2}
  $
  , 
  $  
    \FlaOneByTwo{ U_L } { U_R }
    \leftarrow
    \FlaOneByThreeL{U_0} {U_1} {U_2}
  $
}


\renewcommand{\update}{
  $
  \hspace{1em} 
  \begin{array}{lcl}
    \multicolumn{3}{l}{// \text{Apply transformations from the right.}} \\
    G & \defeq & 
      \mbox{\sc generate\_iid\_stdnorm\_matrix}
                 (m(A) - m(A_{00}), n_b) \\
    Y & \defeq & 
      \left( 
        \FlaTwoByTwoSingleLine{A_{11}}{A_{12}} {A_{21}}{A_{22}}^*
        \FlaTwoByTwoSingleLine{A_{11}}{A_{12}} {A_{21}}{A_{22}} 
      \right)^q
      \FlaTwoByTwoSingleLine{A_{11}}{A_{12}} {A_{21}}{A_{22}}^*
      G \\
    {[}Y,T_V,W_V] & \defeq & 
      \mbox{\sc unpivoted\_QR}(Y) \\
    \FlaTwoByTwoSingleLine{A_{11}}{A_{12}} {A_{21}}{A_{22}} & \defeq &
      \FlaTwoByTwoSingleLine{A_{11}}{A_{12}} {A_{21}}{A_{22}} - 
      \FlaTwoByTwoSingleLine{A_{11}}{A_{12}}{A_{21}}{A_{22}} 
      W_V T_V W_V^* \\
    \FlaOneByTwoSingleLine {V_1}{V_2} & \defeq &
      \FlaOneByTwoSingleLine {V_1}{V_2} - 
      \FlaOneByTwoSingleLine {V_1}{V_2} W_V T_V W_V^* \\
    \multicolumn{3}{l}{// \text{Apply transformations from the left.}} \\
    {[}\FlaTwoByOneSingleLine{A_{11}} {A_{21}},T_U,W_V{]} & \defeq & 
       \mbox{\sc unpivoted\_QR}
       \left( \FlaTwoByOneSingleLine {A_{11}} {A_{21}} \right ) \\
    \FlaTwoByOneSingleLine{A_{12}}{A_{22}} & \defeq & 
      \FlaTwoByOneSingleLine{A_{12}}{A_{22}} - 
      W_U^*T_U^*W_U \FlaTwoByOneSingleLine {A_{12}} {A_{22}} \\
    \FlaOneByTwoSingleLine{U_1}{U_2} & \defeq &
      \FlaOneByTwoSingleLine{U_1}{U_2} - 
      \FlaOneByTwoSingleLine {U_1}{U_2} W_U T_U W_U^* \\
    \multicolumn{3}{l}{// \text{Compute small SVD and update matrices $A$, $U$, and $V$.}} \\
    {[}A_{11}, U_{SVD}, V_{SVD}] & \defeq & \mbox{\sc SVD}(A_{11})\\
    A_{01} & \defeq & A_{01} V_{SVD} \\
    A_{12} & \defeq & U_{SVD}^* A_{12} \\
    V_1 & \defeq & V_1 V_{SVD} \\
    U_1 & \defeq & U_1 U_{SVD}
  \end{array}
  $
}

\begin{figure}[ht!]
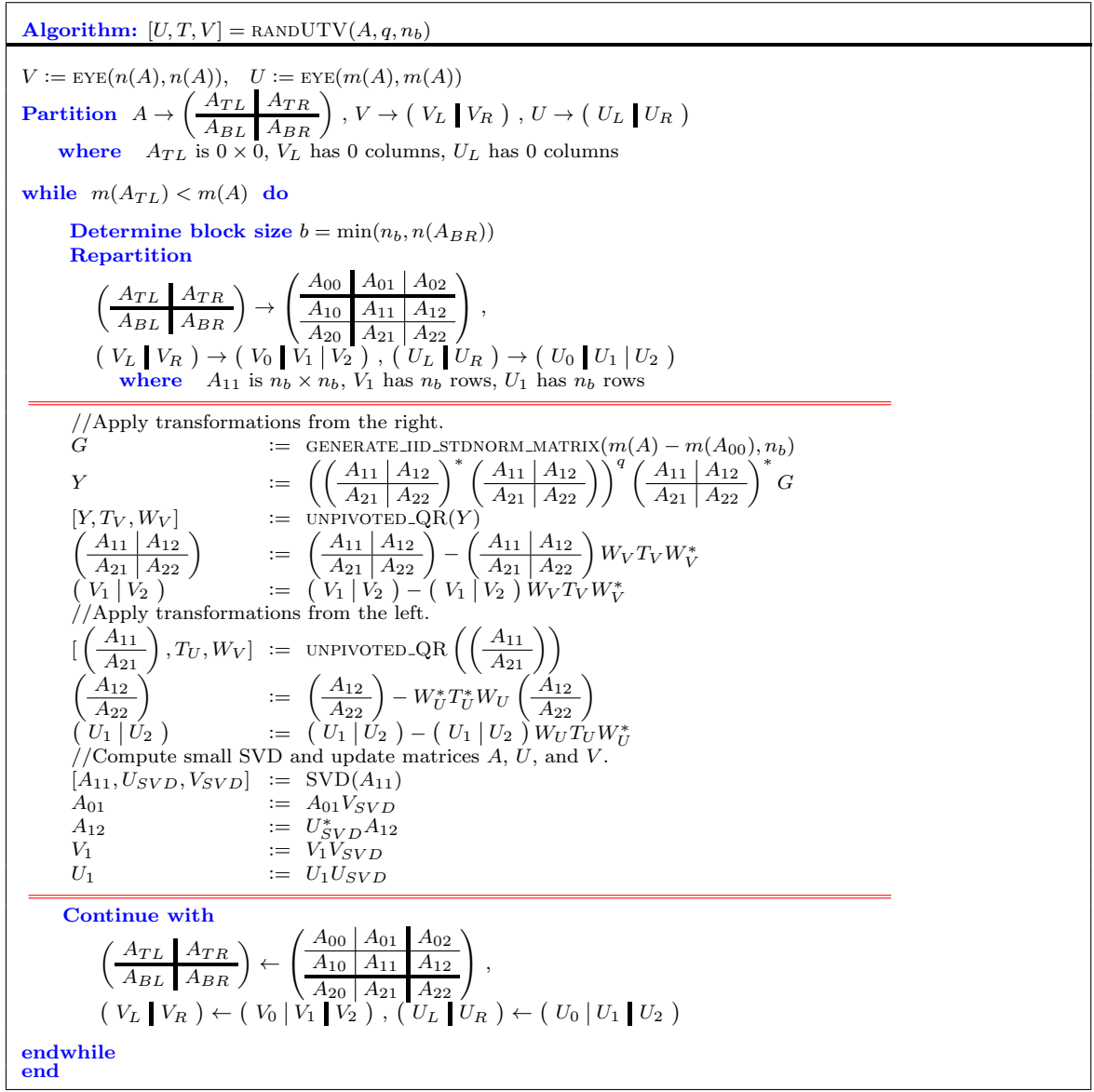

\centering
\begin{minipage}{0.95\textwidth}
\footnotesize
\FlaAlgorithmWithInit
\end{minipage}
\vspace*{-0.2cm}
\caption{The randUTV algorithm
written with the FLAME methodology/notation.  
}
\label{fig:alg_utv}
\end{figure}

\endgroup

\begin{remark}[Connection to randomized SVD]
The randUTV algorithm can be viewed as a blocked and incremental version 
of the randomized SVD (RSVD)  algorithm of \cite{2007_martinsson_PNAS,2011_martinsson_random1,2011_martinsson_randomsurvey}. 
To be precise, the matrix
$$
U(:,1:n_b)\,T(1:n_b,1:n_b)\,(V(:,1:n_b))^{*}
$$
is exactly the rank $n_b$ approximation to $A$ generated by the RSVD, 
as described in \cite[Sec.~1.6]{2011_martinsson_randomsurvey} 
(without ``over sampling'').
\end{remark}

Our description of randUTV was purposefully kept brief in this section, as it has been 
described in detail in a sequence of recent papers. 
The original paper \cite{martinsson2017randutv} contains a detailed discussion of its 
rank revealing properties, with numerical experiments that demonstrate that in practice 
it gets very close to the optimality of the SVD.
Details on implementing it in different computational environments can be found in
\cite{2023_martinsson_randUTV_ooc,2023_martinsson_randUTV_gpu,2022_martinsson_randUTV_parallel}.

\subsection{Adapting randUTV to solve the least squares problem}

When the matrix $A$ does not possess a nullspace, 
the factorization in Eq.~\ref{eq:UTVdef} can be used directly 
to compute the least squares solution of $Ax=b$ through the formula
\begin{equation}
\label{eq:randUTV_solnformula}
x = V\,T(1:n,1:n)^{-1}\,U(1:n,:)^{*}b.
\end{equation}
In the cases where there is a non-trivial nullspace, 
some additional work may be required to ensure that 
both conditions for a least squares solution are satisfied. 
In this case,
the computed UTV factorization takes the form
\begin{equation}
\label{eq:UTVtrunc}
\begin{array}{ccccccccccc}
A &=& U_{1} & [T_{11},\,T_{12}] & V^{*} & ,\\
m\times n && m\times r & r\times n & n\times n
\end{array}
\end{equation}
where $r$ is the rank of the matrix and 
$T_{11}\in \mathbb{R}^{r\times r}$ is upper triangular. 
(In practice, the rank will typically be estimated 
by considering diagonal entries of $T$ below some given threshold $\tau$ to be zero, 
in which case (\ref{eq:UTVtrunc}) holds only approximately.) 
One could now immediately build the solution
\begin{equation}
\label{eq:simplesoln}
x_{\rm simple} = V(:,1:r)\,T_{11}^{-1}\,U_{1}^{*}b ,
\end{equation}
which would be guaranteed to satisfy the first condition of least squares solution
(i.e.~that $\|Ax_{\rm simple} - b\| = \inf_{y}\|Ay-b\|$).
However, $x_{\rm simple}$ is not necessarily the
minimal norm solution among the minimizers. 
To ensure that both conditions are satisfied, 
our software performs by default an additional orthogonal factorization 
of the matrix $[T_{11},\,T_{12}]$, 
as described in Section~\ref{subsec:COD}, and 
conforming to the xGELSY functions in LAPACK~\cite{qqp98}.

Figure~\ref{fig:alg_nullify_t12} shows the algorithm for nullifying $T_{12}$ 
written with the FLAME methodology/notation.
The input arguments are the following:
$T$ is the upper triangular factor of the randUTV factorization of $A$,
$V$ is the right orthogonal matrix of the randUTV factorization, and
$r$ is the numeric rank of $T$.
Matrices $C$, $D$, $E$, and $F$ are views of the original matrices:
Changes made to them are also made to the original matrices.

\begingroup

\setlength{\arraycolsep}{3pt}
\resetsteps 


\renewcommand{\routinename}{ 
  \left[ T, V \right] = \mbox{\sc Nullify\_top\_right\_part\_of\_T}( T, V, r ) 
}


\renewcommand{\initialize}{
  $
    \left( 
      \begin{array}{c|c}
      C & D \\ \hline
      * & * \\
      \end{array}
    \right)
    \defeq T,
  $
    where 
  $ C $ is $ r \times r $,
  $ D $ is $ r \times ( n - r ) $, \\
  $
    \left( 
      \begin{array}{c|c}
      E & F \\
      \end{array}
    \right)
    \defeq V,
  $
    where 
  $ E $ is $ n \times r $,
  $ F $ is $ n \times ( n - r ) $
}


\renewcommand{\partitionings}{
  $
  C \rightarrow \FlaTwoByTwo{C_{TL}}{C_{TR}}
                            {C_{BL}}{C_{BR}}
  $ ,
  $
  D \rightarrow \FlaTwoByOne{D_{T}}
                            {D_{B}}
  $ ,
  $
  E \rightarrow \FlaOneByTwo{E_L}{E_R}
  $
}

\renewcommand{\partitionsizes}{
  $ C_{BR} $ is $ 0 \times 0 $,
  $ D_{B} $ has $ 0 $ rows,
  $ E_R $ has $ 0 $ columns
}


\renewcommand{\guard}{
  m( C_{BR} ) < m( C )
}



\renewcommand{\repartitionings}{
$ \FlaTwoByTwo{C_{TL}}{C_{TR}}
              {C_{BL}}{C_{BR}}
  \rightarrow
  \FlaThreeByThreeTL{C_{00}}{C_{01}}{C_{02}}
                    {C_{10}}{C_{11}}{C_{12}}
                    {C_{20}}{C_{21}}{C_{22}}
$,
$ \FlaTwoByOne{ D_T }
              { D_B }
  \rightarrow
  \FlaThreeByOneT{D_0}
                 {D_1}
                 {D_2}
$, \\
$ \FlaOneByTwo{E_L}{E_R}
  \rightarrow  
  \FlaOneByThreeL{E_0}{E_1}{E_2}
$
}

\renewcommand{\repartitionsizes}{
  $ C_{11} $ is $ b \times b $,
  $ D_1 $ has $ b $ rows,
  $ E_1 $ has $b$ columns
}


\renewcommand{\moveboundaries}{
$ \FlaTwoByTwo{C_{TL}}{C_{TR}}
              {C_{BL}}{C_{BR}}
  \leftarrow
  \FlaThreeByThreeBR{C_{00}}{C_{01}}{C_{02}}
                    {C_{10}}{C_{11}}{C_{12}}
                    {C_{20}}{C_{21}}{C_{22}}
$,
$ \FlaTwoByOne{ D_T }
               { D_B }
  \leftarrow
  \FlaThreeByOneB{D_0}
                 {D_1}
                 {D_2}
$, \\
$ \FlaOneByTwo{E_L}{E_R}
  \leftarrow  
  \FlaOneByThreeR{E_0}{E_1}{E_2}
$
}


\renewcommand{\update}{
  $
  \hspace{1em} 
  \begin{array}{lcl}
    {[}C_{11}, D_{1}] & \defeq & \mbox{\sc Nullify}(C_{11}, D_{1})\\
    {[}C_{01}, D_{0}] & \defeq & \mbox{\sc Update}(C_{11}, D_{1}, C_{01}, D_{0})\\
    {[}E_{1}, F] & \defeq & \mbox{\sc Update}(C_{11}, D_{1}, E_{1}, F)\\
  \end{array}
  $
}

\begin{figure}[th!]
\centering
\begin{minipage}{0.95\textwidth}
\footnotesize
\FlaAlgorithmWithInit
\end{minipage}
\vspace*{-0.2cm}
\caption{The algorithm to nullify the top right part of $T$
written with the FLAME methodology/notation.
The input arguments are the following:
$T$ is the upper triangular factor of the randUTV factorization of $A$,
$V$ is the right orthogonal matrix of the randUTV factorization, and
$r$ is the numeric rank of $T$.
}
\label{fig:alg_nullify_t12}
\end{figure}

\endgroup

Figure~\ref{fig:alg_axb} shows the overall algorithm algorithm 
for solving $Ax=b$ written with the FLAME methodology/notation.

\begingroup

\setlength{\arraycolsep}{3pt}
\resetsteps 


\renewcommand{\routinename}{ 
  \left[ x \right] = \mbox{\sc Solve\_linear\_system}( A, b, q )
}


\renewcommand{\update}{
  $
  \hspace{1em} 
  \begin{array}{lcl}
  {[}U, T, V] & \defeq & \mbox{\sc randUTV}(A, q, n_b) \\
  {[}r]       & \defeq & \mbox{\sc Compute\_rank}(T) \\
  {[}T, V]    & \defeq & \mbox{\sc Nullify\_top\_right\_part\_of\_T}(T, V, r) \\
  {[}x]       & \defeq & V ( T_{tl}^{-1} ( U^* b ) ),
                         \mbox{ where } T_{tl} = T( 1:r, 1:r ) \\
  \end{array}
  $
}

\begin{figure}[th!]
\centering
\begin{minipage}{0.95\textwidth}
\footnotesize
\FlaAlgorithmWithoutLoop
\end{minipage}
\vspace*{-0.2cm}
\caption{The overall algorithm for solving $Ax=b$.
The input arguments are the following:
$A$ is the coefficient matrix of dimension $m \times n$,
$b$ is the independent vector of dimension $m \times 1$, and
$q$ is the number of steps in the power iteration process.
}
\label{fig:alg_axb}
\end{figure}

\endgroup

\begin{remark}[Fast option]
Due to the fact that randUTV is very precise in revealing the numerical rank, 
the extra orthonormalization step can in fact be dropped with minimal impact
on the norm of the residual. Our software offers this ``accelerated version''
as an option for users who do not need a guarantee that the absolute minimal
norm solution is returned. 

\end{remark}

\subsection{Blocked implementation for In-Core data}
\label{subsec:trad}

When solving linear systems with orthogonal transformations,
the most-common computational tasks are 
the computation of a Householder transformation and 
its application to some matrix columns (or rows).
Scalar algorithms apply these transformations one by one 
to the rest of the matrix (right-looking algorithms) or 
to the current column (left-looking algorithms).
To apply one Householder transformation to the rest of the matrix, 
all the rest of the matrix must be read into the CPU,
which requires a lot of memory accesses.

In contrast,
blocked algorithms apply several Householder transformations 
at the same time, 
thus reducing the amount of data being transferred
between main memory and the CPU.
Moreover, they allow the use of highly-efficient matrix-matrix operations,
which perform many floating-point operations per memory access 
(high arithmetic intensity), which is key to obtain high performances.

Figure~\ref{fig:qr_blocked} shows a graphical representation
of a blocked algorithm for computing the QR factorization 
with a right-looking algorithm (others factorizations would be similar).
For the sake of brevity, this figure shows only its first tasks.
In this figure,
the `$\star$' symbol represents an element modified by the current task.
The `$\circ$' symbol represents an element that has been nullified.
Note that `$\circ$' elements do not actually store zeros but information 
about the Householder transformations that will be later used to apply them.
The continuous lines surround the blocks involved (read or written)
in the current task.


\begingroup
\setlength{\tabcolsep}{3.0pt}       
\renewcommand{\arraystretch}{1.00}  

\newcommand{\newvs}{\vspace*{0.4cm}}
\newcommand{\newminusvs}{\vspace*{-0.1cm}}

\newcommand{\mybl}{$\bullet$}
\newcommand{\mybn}{\multicolumn{1}{c}{$\bullet$}} 
\newcommand{\mysl}{$\star$}
\newcommand{\mysn}{\multicolumn{1}{c}{$\star$}}
\newcommand{\mypl}{$\circ$}
\newcommand{\mypn}{\multicolumn{1}{c}{$\circ$}}

\renewcommand{\thesubfigure}{\arabic{subfigure}}

\newcolumntype{C}{>{\centering\arraybackslash}p{4.0pt}}

\begin{figure}
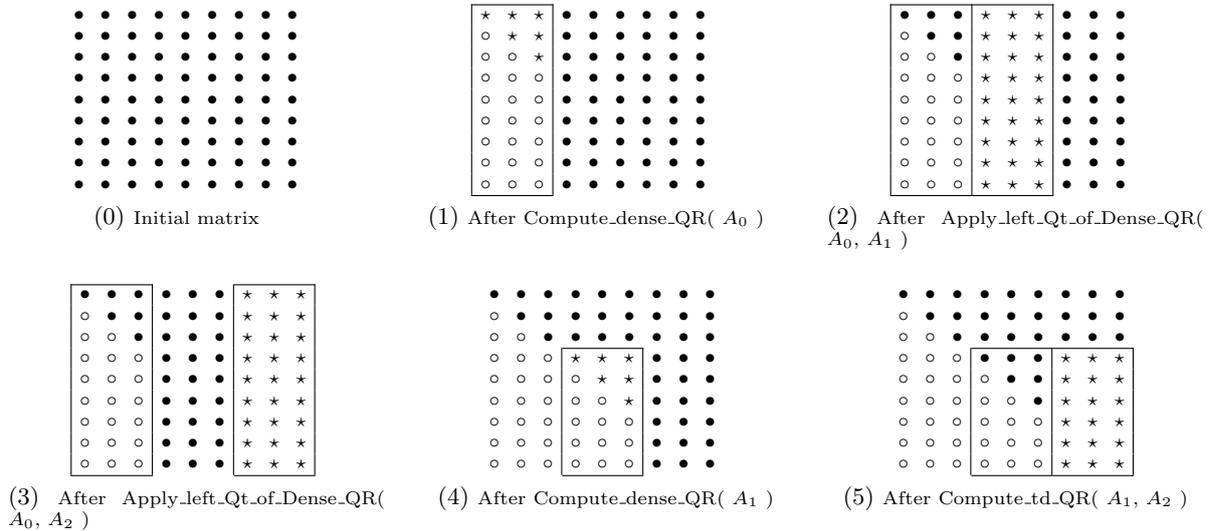

\begin{subfigure}[t]{0.31\linewidth}
\hspace*{0.1cm}
%
\setcounter{subfigure}{-1}
\scriptsize
\centering
\begin{tabular}{CCCCCCCCC} 
  \mybl & \mybl & \mybl & \mybl & \mybl & \mybl & \mybl & \mybl & \mybl \\
  \mybl & \mybl & \mybl & \mybl & \mybl & \mybl & \mybl & \mybl & \mybl \\
  \mybl & \mybl & \mybl & \mybl & \mybl & \mybl & \mybl & \mybl & \mybl \\ 
  \mybl & \mybl & \mybl & \mybl & \mybl & \mybl & \mybl & \mybl & \mybl \\
  \mybl & \mybl & \mybl & \mybl & \mybl & \mybl & \mybl & \mybl & \mybl \\
  \mybl & \mybl & \mybl & \mybl & \mybl & \mybl & \mybl & \mybl & \mybl \\
  \mybl & \mybl & \mybl & \mybl & \mybl & \mybl & \mybl & \mybl & \mybl \\
  \mybl & \mybl & \mybl & \mybl & \mybl & \mybl & \mybl & \mybl & \mybl \\
  \mybl & \mybl & \mybl & \mybl & \mybl & \mybl & \mybl & \mybl & \mybl \\
\end{tabular}
\newminusvs
\caption{\scriptsize Initial matrix}
\newvs
\end{subfigure}
\hspace*{0.1cm}
\begin{subfigure}[t]{0.31\linewidth}
\scriptsize
\centering
\begin{tabular}{|CCC|CCCCCC} 
\cline{1-3}
  \mysl & \mysl & \mysl & \mybl & \mybl & \mybl & \mybl & \mybl & \mybl \\
  \mypl & \mysl & \mysl & \mybl & \mybl & \mybl & \mybl & \mybl & \mybl \\
  \mypl & \mypl & \mysl & \mybl & \mybl & \mybl & \mybl & \mybl & \mybl \\ 
  \mypl & \mypl & \mypl & \mybl & \mybl & \mybl & \mybl & \mybl & \mybl \\
  \mypl & \mypl & \mypl & \mybl & \mybl & \mybl & \mybl & \mybl & \mybl \\
  \mypl & \mypl & \mypl & \mybl & \mybl & \mybl & \mybl & \mybl & \mybl \\
  \mypl & \mypl & \mypl & \mybl & \mybl & \mybl & \mybl & \mybl & \mybl \\
  \mypl & \mypl & \mypl & \mybl & \mybl & \mybl & \mybl & \mybl & \mybl \\
  \mypl & \mypl & \mypl & \mybl & \mybl & \mybl & \mybl & \mybl & \mybl \\
\cline{1-3}
\end{tabular}
\newminusvs
\caption{\scriptsize After Compute\_dense\_QR( $A_{0}$ )}
\newvs
\end{subfigure}
\hspace*{0.1cm}
\begin{subfigure}[t]{0.31\linewidth}
\scriptsize
\centering
\begin{tabular}{|CCC|CCC|CCC} 
\cline{1-6}
  \mybl & \mybl & \mybl & \mysl & \mysl & \mysl & \mybl & \mybl & \mybl \\
  \mypl & \mybl & \mybl & \mysl & \mysl & \mysl & \mybl & \mybl & \mybl \\
  \mypl & \mypl & \mybl & \mysl & \mysl & \mysl & \mybl & \mybl & \mybl \\ 
  \mypl & \mypl & \mypl & \mysl & \mysl & \mysl & \mybl & \mybl & \mybl \\
  \mypl & \mypl & \mypl & \mysl & \mysl & \mysl & \mybl & \mybl & \mybl \\
  \mypl & \mypl & \mypl & \mysl & \mysl & \mysl & \mybl & \mybl & \mybl \\
  \mypl & \mypl & \mypl & \mysl & \mysl & \mysl & \mybl & \mybl & \mybl \\
  \mypl & \mypl & \mypl & \mysl & \mysl & \mysl & \mybl & \mybl & \mybl \\
  \mypl & \mypl & \mypl & \mysl & \mysl & \mysl & \mybl & \mybl & \mybl \\
\cline{1-6}
\end{tabular}
\newminusvs
\caption{\scriptsize After Apply\_left\_Qt\_of\_Den\-se\_QR( $A_{0}$, $A_{1}$ )}
\newvs
\end{subfigure}
\hspace*{0.1cm}
\begin{subfigure}[t]{0.31\linewidth}
\scriptsize
\centering
\begin{tabular}{|CCC|CCC|CCC|} 
\cline{1-3}\cline{7-9}
  \mybl & \mybl & \mybl & \mybl & \mybl & \mybl & \mysl & \mysl & \mysl \\
  \mypl & \mybl & \mybl & \mybl & \mybl & \mybl & \mysl & \mysl & \mysl \\
  \mypl & \mypl & \mybl & \mybl & \mybl & \mybl & \mysl & \mysl & \mysl \\ 
  \mypl & \mypl & \mypl & \mybl & \mybl & \mybl & \mysl & \mysl & \mysl \\
  \mypl & \mypl & \mypl & \mybl & \mybl & \mybl & \mysl & \mysl & \mysl \\
  \mypl & \mypl & \mypl & \mybl & \mybl & \mybl & \mysl & \mysl & \mysl \\
  \mypl & \mypl & \mypl & \mybl & \mybl & \mybl & \mysl & \mysl & \mysl \\
  \mypl & \mypl & \mypl & \mybl & \mybl & \mybl & \mysl & \mysl & \mysl \\
  \mypl & \mypl & \mypl & \mybl & \mybl & \mybl & \mysl & \mysl & \mysl \\
\cline{1-3}\cline{7-9}
\end{tabular}
\newminusvs
\caption{\scriptsize After Apply\_left\_Qt\_of\_Den\-se\_QR( $A_{0}$, $A_{2}$ )}
\newvs
\end{subfigure}
\hspace*{0.1cm}
\begin{subfigure}[t]{0.31\linewidth}
\scriptsize
\centering
\begin{tabular}{CCC|CCC|CCC} 
  \mybl & \mybl & \mybn & \mybl & \mybl & \mybn & \mybl & \mybl & \mybl \\
  \mypl & \mybl & \mybn & \mybl & \mybl & \mybn & \mybl & \mybl & \mybl \\
  \mypl & \mypl & \mybn & \mybl & \mybl & \mybn & \mybl & \mybl & \mybl \\ 
\cline{4-6}
  \mypl & \mypl & \mypl & \mysl & \mysl & \mysl & \mybl & \mybl & \mybn \\
  \mypl & \mypl & \mypl & \mypl & \mysl & \mysl & \mybl & \mybl & \mybn \\
  \mypl & \mypl & \mypl & \mypl & \mypl & \mysl & \mybl & \mybl & \mybn \\
  \mypl & \mypl & \mypl & \mypl & \mypl & \mypl & \mybl & \mybl & \mybn \\
  \mypl & \mypl & \mypl & \mypl & \mypl & \mypl & \mybl & \mybl & \mybn \\
  \mypl & \mypl & \mypl & \mypl & \mypl & \mypl & \mybl & \mybl & \mybn \\
\cline{4-6}
\end{tabular}
\newminusvs
\caption{\scriptsize After Compute\_dense\_QR( $A_{1}$ )}
\newvs
\end{subfigure}
\hspace*{0.1cm}
\begin{subfigure}[t]{0.31\linewidth}
\scriptsize
\centering
\begin{tabular}{CCC|CCC|CCC|} 
  \mybl & \mybl & \mybn & \mybl & \mybl & \mybn & \mybl & \mybl & \mybn \\
  \mypl & \mybl & \mybn & \mybl & \mybl & \mybn & \mybl & \mybl & \mybn \\
  \mypl & \mypl & \mybn & \mybl & \mybl & \mybn & \mybl & \mybl & \mybn \\ 
\cline{4-6}\cline{7-9}
  \mypl & \mypl & \mypl & \mybl & \mybl & \mybl & \mysl & \mysl & \mysl \\
  \mypl & \mypl & \mypl & \mypl & \mybl & \mybl & \mysl & \mysl & \mysl \\
  \mypl & \mypl & \mypl & \mypl & \mypl & \mybl & \mysl & \mysl & \mysl \\
  \mypl & \mypl & \mypl & \mypl & \mypl & \mypl & \mysl & \mysl & \mysl \\
  \mypl & \mypl & \mypl & \mypl & \mypl & \mypl & \mysl & \mysl & \mysl \\
  \mypl & \mypl & \mypl & \mypl & \mypl & \mypl & \mysl & \mysl & \mysl \\
\cline{4-6}\cline{7-9}
\end{tabular}
\newminusvs
\caption{\scriptsize After Compute\_td\_\-QR( $A_{1}$, $A_{2}$ )}
\newvs
\end{subfigure}
\vspace*{-0.4cm}
\caption{An illustration of the first tasks performed by a blocked algorithm
for computing the QR factorization.
The `$\bullet$' symbol represents an element not modified by the current task,
`$\star$' represents an element modified by the current task, and
`$\circ$' represents a nullified element.
The continuous lines surround the blocks involved in the current task.}
\label{fig:qr_blocked}
\end{figure}

\endgroup


Martinsson \textit{et al.}~\cite{martinsson2017randutv}
developed a blocked algorithm for computing the randUTV.
The experimental study showed that 
their implementation was competitive with high-performance implementations
of both the SVD and CPQR from the Intel MKL library.
Although the randUTV requires many more floating-point operations
than the other two factorizations (CPQR and SVD),
most of those floating-point operations are done 
inside BLAS-3 subroutines (matrix-matrix operations),
which yields a high computational speed.

\subsection{Algorithm-by-blocks implementation for In-Core data}
\label{subsec:abb}

Although obtaining higher performances than scalar algorithms,
blocked algorithms usually contain a performance bottleneck.
The applying of the transformations to the rest of matrix 
is usually very efficient:
BLAS-3 operations can be applied 
and all cores are usually busy doing work.
On the other side,
the computation of the transformations 
usually involve only a small number of columns 
with only a small amount of floating-point operations, and
during these, only one core is busy and the rest of cores are idle, 
which can greatly decrease performances when having many cores.

To overcome this problem,
algorithms-by-blocks work in a different way.
They apply the traditional algorithm
on square blocks instead of scalars,
and hence the name.
The original scalar algorithm is converted on a series of tasks,
where every task operates on a few blocks 
(usually two, three, or four).
Figure~\ref{fig:qr_ab} shows a graphical representation
of a right-looking variant of 
an algorithm-by-blocks for computing a QR factorization.
For the sake of brevity,
this figure shows only its first tasks.
In this figure,
blocks involved in the current task are in bold face,
blocks modified by the current task are marked with the `$^*$' superscript,
and blocks nullified are marked with the `$0$' symbol.
Note that nullified blocks do not actually store zeros but information 
about the Householder transformations that will be later used to apply them.


\begingroup

\newcommand{\newvs}{\vspace*{0.4cm}}
\newcommand{\newminusvs}{\vspace*{-0.1cm}}

\renewcommand{\thesubfigure}{\arabic{subfigure}}

\begin{figure}[ht!]

\begin{subfigure}[t]{.30\linewidth}
\setcounter{subfigure}{-1}
\centering
\(
  \left(
  \begin{array}{c|c|c}
    A_{00}   & A_{01}   & A_{02}   \\ \hline
    A_{10}   & A_{11}   & A_{12}   \\ \hline
    A_{20}   & A_{21}   & A_{22}   \\
  \end{array}
  \right)
\)
\newminusvs
\caption{\footnotesize Initial matrix}
\newvs
\end{subfigure}
\hspace*{0.3cm}
\begin{subfigure}[t]{.30\linewidth}
\centering
\(
  \left(
  \begin{array}{c|c|c}
    \mathbf{A_{00}^*} & A_{01}            & A_{02}   \\ \hline
    A_{10}            & A_{11}            & A_{12}   \\ \hline
    A_{20}            & A_{21}            & A_{22}   \\
  \end{array}
  \right)
\)
\newminusvs
\caption{\footnotesize After Compute\_dense\_QR( $A_{00}$ )}
\newvs
\end{subfigure}
\hspace*{0.3cm}
\begin{subfigure}[t]{.30\linewidth}
\centering
\(
  \left(
  \begin{array}{c|c|c}
    \mathbf{A_{00}}   & \mathbf{A_{01}^*} & A_{02}   \\ \hline
    A_{10}            & A_{11}            & A_{12}   \\ \hline
    A_{20}            & A_{21}            & A_{22}   \\
  \end{array}
  \right)
\)
\newminusvs
\caption{\footnotesize After Apply\_left\_Qt\_of\_Den\-se\_QR( $A_{00}$, $A_{01}$ )}
\newvs
\end{subfigure}
\hspace*{0.3cm}
\begin{subfigure}[t]{.30\linewidth}
\centering
\(
  \left(
  \begin{array}{c|c|c}
    \mathbf{A_{00}}   & A_{01}            & \mathbf{A_{02}^*} \\ \hline
    A_{10}            & A_{11}            & A_{12}   \\ \hline
    A_{20}            & A_{21}            & A_{22}   \\
  \end{array}
  \right)
\)
\newminusvs
\caption{\footnotesize After Apply\_left\_Qt\_of\_Den\-se\_QR( $A_{00}$, $A_{02}$ )}
\newvs
\end{subfigure}
\hspace*{0.3cm}
\begin{subfigure}[t]{.30\linewidth}
\centering
\(
  \left(
  \begin{array}{c|c|c}
    \mathbf{A_{00}^*} & A_{01}            & A_{02}   \\ \hline
    \mathbf{0     ^*} & A_{11}            & A_{12}   \\ \hline
    A_{20}            & A_{21}            & A_{22}   \\
  \end{array}
  \right)
\)
\newminusvs
\caption{\footnotesize After Compute\_td\_\-QR( $A_{00}$, $A_{10}$ )}
\newvs
\end{subfigure}
\hspace*{0.3cm}
\begin{subfigure}[t]{.30\linewidth}
\centering
\(
  \left(
  \begin{array}{c|c|c}
    \mathbf{A_{00}}   & \mathbf{A_{01}^*} & A_{02}   \\ \hline
    \mathbf{0}        & \mathbf{A_{11}^*} & A_{12}   \\ \hline
    A_{20}            & A_{21}            & A_{22}   \\
  \end{array}
  \right)
\)
\newminusvs
\caption{\footnotesize After Apply\_left\_Qt\_of\_\-td\_QR( $A_{00}$, $A_{10}$, $A_{01}$, $A_{11}$ )}
\newvs
\end{subfigure}
\hspace*{0.3cm}
\begin{subfigure}[t]{.30\linewidth}
\centering
\(
  \left(
  \begin{array}{c|c|c}
    \mathbf{A_{00}}   & A_{01}            & \mathbf{A_{02}^*} \\ \hline
    \mathbf{0}        & A_{11}            & \mathbf{A_{12}^*} \\ \hline
    A_{20}            & A_{21}            & A_{22}   \\
  \end{array}
  \right)
\)
\newminusvs
\caption{\footnotesize After Apply\_left\_Qt\_of\_\-td\_QR( $A_{00}$, $A_{10}$, $A_{02}$, $A_{12}$ )}
\newvs
\end{subfigure}
\hspace*{0.3cm}
\begin{subfigure}[t]{.30\linewidth}
\centering
\(
  \left(
  \begin{array}{c|c|c}
    \mathbf{A_{00}^*} & A_{01}            & A_{02}   \\ \hline
    0                 & A_{11}            & A_{12}   \\ \hline
    \mathbf{0     ^*} & A_{21}            & A_{22}   \\
  \end{array}
  \right)
\)
\newminusvs
\caption{\footnotesize After Compute\_td\_\-QR( $A_{00}$, $A_{20}$ )}
\newvs
\end{subfigure}
\hspace*{0.3cm}
\begin{subfigure}[t]{.30\linewidth}
\centering
\(
  \left(
  \begin{array}{c|c|c}
    \mathbf{A_{00}}   & \mathbf{A_{01}^*} & A_{02}   \\ \hline
    0                 & A_{11}            & A_{12}   \\ \hline
    \mathbf{0}        & \mathbf{A_{21}^*} & A_{22}   \\
  \end{array}
  \right)
\)
\newminusvs
\caption{\footnotesize After Apply\_left\_Qt\_of\_\-td\_QR( $A_{00}$, $A_{20}$, $A_{01}$, $A_{21}$ )}
\newvs
\end{subfigure}
\hspace*{0.3cm}
\vspace*{-0.4cm}
\caption{An illustration of the first tasks performed 
by an algorithm-by-blocks for computing the QR factorization. 
Blocks involved in the current task are in bold face,
blocks modified by the current task are marked with the `$^*$' superscript,
and blocks nullified are marked with the `$0$' symbol.
}
\label{fig:qr_ab}
\end{figure}

\endgroup

Like blocked algorithms,
algorithms-by-blocks also reduce the amount of data transferred
and allow to use BLAS-3 subroutines.
In this case, BLAS-3 subroutines are used inside each task.
Another advantage is that many tasks on blocks 
can be performed in parallel.
For instance, 
in the previous figure, 
tasks in subfigures 2 and 3 can be simultaneously executed.
Moreover,
tasks in subfigures 2, 3, and 4 could be simultaneously executed 
if a copy of the Householder transformations in $A_{00}$ was done before.
The main advantage of algorithm-by-blocks is that 
they do not contain parts with strong bottlenecks
in which only one core can work and the remaining ones are idle.
In this type of algorithms, several tasks can be usually performed in parallel.
And even if one task could not be overlapped, 
its computational cost would be small 
(since all the operands are $n_b \times n_b$),
and hence the bottleneck would be small too.
In contrast, blocked algorithms usually have larger bottlenecks.
For instance, in the blocked column-oriented right-looking algorithm 
for computing the QR factorization,
the bottleneck is the factorization of the current column block,
which can be up to $m \times n_b$.

Heavner \textit{et al.}~\cite{2022_martinsson_randUTV_parallel}
developed and assessed an algorithm-by-blocks 
for computing the randUTV factorization 
on shared-memory multicore and multiprocessor architectures.
Their implementations of the algorithm-by-blocks 
for computing the randUTV factorization can be 
more than twice as fast as the implementations of the blocked algorithms
for the same computation.
That work also shows that despite performing 
between 3.5 ($q=0$) about 5.5 ($q=2$)
as many floating-point operations as the CPQR, 
it is several times faster than it
(because of being very rich in BLAS-3 operations, as previously told).
Moreover, it is also faster than a very high-performance implementation 
of the SVD factorization included in Intel MKL.

\subsection{Out-Of-Core implementations 
            for the least-squares problem with the randUTV}
\label{sec:abb-randutv}

In some problems, the data to be processed is much larger
than the available main memory in the computer.
In these cases, 
there are several options.
The first one is to acquire a computer with a main memory 
so large that the data fits inside it.
The second one is
to use a distributed-memory architecture
with multiple nodes so that the total main memory is larger,
which requires several nodes (computers) and 
a high-speed communication network to connect them.
In either case,
the cost is much higher and in some cases it could be beyond affordable.

A different option is to rewrite the software 
so that not all of the data is stored in main memory all of the time.
By rewriting the software,
affordable computers already acquired or 
newly acquired affordable computers can be employed.

The traditional technique is to decompose the work to do 
into several computational major tasks.
Then, for every task, the data to be used must be loaded
from secondary storage (disk) to main memory,
then processed there, and finally the data modified must be copied back
from main memory to secondary storage.
Blocked algorithms can employ this pattern.
For instance, 
in a right-looking algorithm for computing the QR factorization,
a block of columns is loaded, then factorized, and 
finally written back into disk storage.
Then, the rest of the matrix must be updated also by blocks of columns.
That is, a block of columns of the rest of the matrix is loaded, 
then modified, and then written back into disk.
See Figure~\ref{fig:qr_blocked} again for a graphical representation
of a blocked algorithm for computing a factorization.

However, as can be easily seen in this figure, 
when applying this technique,
the amount of data being transferred and processed is smaller in each iteration.
For instance, 
compare the amount of data transferred and processed 
in the first iteration (subfigures 1, 2, and 3) and 
in the second iteration (subfigures 4 and 5).
The problem with this approach is that
after each iteration of the blocked algorithm,
main memory is taken advantage less and less.
In the final iterations of the blocked algorithm,
only a small amount of the main memory is used.
If we want to take advantage of most of main memory in the first
iteration, we must adjust the block size 
according to the main memory size and to the number of rows $m$.
And that only guarantees maximum usage of main memory 
in the first iteration.

In contrast,
algorithm-by-blocks can also match 
the pattern of loading the data to be used 
from secondary storage (disk) to main memory,
then process it, and finally copy back the data modified
from main memory into secondary storage.
In this case, 
in order to take full advantage of main memory,
the block size does not depend on $m$ and $n$ 
(the dimensions of matrix $A$) at all.
The amount of data transferred and processed 
does not decrease as the algorithm iteration advances,
and it depends only on the type of task being executed.


Heavner \textit{et al.}~\cite{2023_martinsson_randUTV_ooc}
developed an Out-Of-Core implementation of the randUTV factorization 
attaining high performance
for CPU architectures.

In our work,
we have applied some performance improvements to this previous implementation
of the randUTV factorization for CPU architectures:
The use of broader (more complex) computational kernels and 
the implementation of a new block cache management policy.
Moreover, we have developed a new family of implementations for GPU architectures.
This implementation is further refined 
to solve the linear least squares problem
by incorporating the additional three tasks:
The first one is a simple rank estimation step.
The second one is the annihilation of the top right block 
of dimension $r \times (n-r)$ of $T$
to provide a clear split 
between the nullspace of the matrix and its orthogonal complement. 
The third task is the final computations to obtain the solution $x$.

\subsubsection{New implementations}

In this work we present several implementations
for solving the linear least squares problem 
with rank-deficient (or full-rank) coefficient matrices
by using the randUTV factorization.
%
For a better performance analysis (by isolating every technique),
each one of the following variants usually contributes 
a specific improvement with respect to the previous one.
The variants for CPU architectures are described next:

\begin{itemize}

\item
\texttt{v21t}:
This is a straightforward implementation for solving
rank-deficient linear systems by using the randUTV factorization
on CPU multicore systems.
In this case both matrices $U$ and $V$ are explicitly built.
This is the traditional Out-Of-Core implementation:
In order to perform every task,
all the input operands are brought from disk to main memory,
then they are processed, and 
finally the output operands are written back to disk.

\item
\texttt{v22t}:
As said, this is a modification of the previous one \texttt{v21t}.
%
Previous variant \texttt{v21t} is implemented with the
traditional approach of using BLAS-3 subroutines 
(\texttt{dgemm}, \texttt{dtrsm}, etc.)
from a high-performance BLAS library such as the Intel MKL library
since it usually renders high performance gains.
In contrast, the \texttt{v22t} variant employs broader (more complex) 
computational methods 
(\texttt{dlarfb}, \texttt{dgeqrf}, \texttt{dormqr}, etc.)
from the same library
since they have been greatly optimized by this company.

\item
\texttt{v23t}:
Modification of \texttt{v22t}.
In this variant, matrix $U$ is not explicitly built,
and the left orthogonal transformations are applied to $b$ (or $B$) on-the-fly.
The main advantage is that a lot of computation and data transfers
can be saved since the full matrix $U$ is not computed, 
thus obtaining higher performances.
In this case, we can solve 
as many linear systems (as many columns in $B$) as desired.
On the other hand, if this approach is used,
the randUTV factorization cannot be later reused to solve 
other linear systems with a different matrix $B$.


\item
\texttt{v23c}:
Modification of \texttt{v23t}.
It employs a block cache 
with the aim of saving transfers of operands between main memory and disk.
For instance, if two consecutive tasks are the following:
\texttt{Compute\_dense\_QR(} $A_{00}$ \texttt{)} and
\texttt{Apply\_dense\_QR(} $A_{00}, A_{01}$ \texttt{)},
then the traditional approach loads operand $A_{00}$, processes it,
and then stores it.
Then, to process the second task, the operand $A_{00}$ must be loaded again.
By using a cache of blocks stored in main memory,
the second load of $A_{00}$ can be saved,
thus reducing the amount of data transfers between main memory and disk.

To accelerate the cache management, 
all slots in the block cache are created with the maximum square block size.
It also employs a four-set associative method
(a block can only be stored in one of the four sets)
to reduce the block search.
The eviction cache policy to discard blocks 
when the cache is full (and no other block can be brought)
is the LRU (Least-Recently Used) policy,
since it usually renders good results.

\item
\texttt{v23d}:
Modification of \texttt{v23c}.
It employs a block cache with the same LRU (Least-Recently Used) policy.
This variant introduces two improvements 
with respect to the previous one:
The first one is that the slots in the block cache 
are of the same size of the blocks actually stored.
The second one is that a block can be stored in any position 
(no four-set associative method).
The first change allows to fit more blocks in the cache, but 
requires a more complex management.
The second change allows a block to be stored in any gap,
but can make the block search slower.

\item
\texttt{v23e}:
Modification of \texttt{v23d}.
The LRU policy for cache replacement
offers usually good results (a good hit/miss ratio)
since it tries the reduce the number of cache misses
by using the past behavior to foresee the next data transfers.
In our case, 
since the list of tasks has been already completely built 
before the execution starts, 
it can effectively be used to choose the block to be replaced.
In this case,
we have decided to use the 
LFU (Latest in the Future to be Used) eviction policy.
When having to discard one block, 
the block chosen to be discarded is
the block being used latest in the future,
which guarantees a very good hit/miss ratio.

\item
\texttt{v23x}:
Modification of \texttt{v23e}.
In this case, 
the overlapping of computation and disk I/O is employed 
to improve performances again.
To achieve this, 
the implementations use two threads:
the disk I/O thread and the computational thread.
The disk I/O thread manages all the I/O of the program and the block cache.
When processing tasks,
it tries to stay ahead of the computational thread 
so that all operands are ready when the computational thread
wants to execute a task.
To do that, the disk I/O thread must bring operands in advance,
and then it must write back operands when the cache is full.
The computational thread executes the tasks if the operands are ready.
If the operands are not ready, the computational thread must wait.
In this case, both threads work according to the producer-consumer pattern.

\item
\texttt{v23s}:
Modification of \texttt{v23x}.
Unlike the previous one,
this variant uses page-locked memory (\textit{pinned} memory)
for the whole block cache.
The page-locked memory is a user-defined part of the main memory that 
the operating system cannot move to disk 
when the operating system has virtual memory 
and there is not enough main memory to store all the programs and data.
This can accelerate some accesses because 
this part of the memory cannot ever be chosen 
to be moved into the disk by the operating system.
In the case of CPU variants, the page-locked memory is asked 
to the operating system with one system call.

\item
\texttt{v24s}:
Modification of \texttt{v23s}.
This variant does not annihilate block $T_{12}$
after computing the randUTV factorization,
since the norm of this block is usually small 
after this factorization. 
Recall that block $T_{12}$ is 
the top right block of the computed factor $T$ 
with dimension $r \times ( n - r )$,
where $r$ and $n$ are the numerical rank and
the number of columns of matrix $A$, respectively.

\end{itemize}

The variants for GPU architectures are described next.
In this case, this number is smaller since we have used 
the insights of CPU variants to reduce them (as well as the number of experiments).
In the case of GPU variants, 
we have used the C programming language
and the cuSOLVER and cuBLAS libraries.

\begin{itemize}

\item
\texttt{v33t}:
Migration of \texttt{v23t} for GPU systems.

\item
\texttt{v33c}:
Migration of \texttt{v23c} for GPU systems.

\item
\texttt{v33d}:
Migration of \texttt{v23d} for GPU systems.

\item
\texttt{v33e}:
Migration of \texttt{v23e} for GPU systems.

\item
\texttt{v33x}:
Migration of \texttt{v23x} for GPU systems.

\item
\texttt{v33s}:
Migration of \texttt{v23s} for GPU systems.
In this variant the page-locked memory is asked to the NVIDIA driver
as performances are better than when asked to the operating system.
The page-locked memory is managed by the operating system,
but asking to the NVIDIA driver helped it to be much more aware of it.

\item
\texttt{v34s}:
Migration of \texttt{v24s} for GPU systems.

\end{itemize}

\section{Numerical analysis}
\label{sec:numerical_analysis}

In this section, we investigate the numerical behavior 
of the new Out-Of-Core implementations 
for solving the linear least-squares problem
and compare it with  current high-performance implementations
that work with data stored in main memory.
First, the precision of the implementations is assessed;
then, the performances are assessed.
In all the experiments, double-precision real matrices were processed.

\subsection{Precision analysis}
\label{subsec:precision_analysis}

This subsection contains an analysis of the precision obtained 
when solving the least-squares problem using different scenarios.

In a preliminary precision analysis we compared all of our new implementations 
for solving the linear least-squares problem.
The results showed that all of them obtained similar results.
For the sake of brevity,
in the rest of this subsection we only report two variants,
\texttt{v33s} and \texttt{v34s},
since their approach is very different:
The first one employs the usual complete method,
whereas the second one skips the annihilation of $T_{12}$.

In all the precision experiments,
we have used $q=0$ (number of steps in the power iteration process),
since this number has been usually enough to reveal the rank with accuracy.
If a good approximation to the singular values were required, 
then this number should be larger 
(usually $q=1$ and $q=2$ deliver good approximations 
of up to several digits to the singular values).

To be used as a reference,
we also assessed two currently widely used implementations
that work on main memory (in-core):

\begin{itemize}

\item 
\texttt{MKL\_GELSS}:
The linear least-squares solver \texttt{dGELSS} from the Intel MKL library.
This implementation is based on the SVD 
(see Subsection~\ref{subsec:SVD}).

\item
\texttt{MKL\_GELSY}:
The linear least-squares solver \texttt{dGELSY} from the Intel MKL library.
This implementation is based on a 
complete orthogonal decomposition 
based on the column-pivoting QR factorization
(see Subsection~\ref{subsec:COD}).

\end{itemize}

The platform employed to assess the precision was \texttt{aurus}.
It is a server with two AMD EPYC 7282 processors (32 cores in total)
at a base clock frequency of 2.8 GHz
and 512 GiB of RAM. 
This server also has an NVIDIA A100 GPU with 80 GiB of RAM. 
The operating system is Rocky Linux 8.7. 
The C compiler is GNU gcc 8.5.0 20210514 (Red Hat 8.5.0-16). 
The computational libraries employed are
Intel oneAPI Math Kernel Library Version 2023.1-Product Build 20230303
and Nvidia CUDA 11.7.

\subsubsection{Test matrices}

In order to assess the precision, a set of test matrices 
from the Suite Sparse Matrix Collection~\cite{davis2011university} 
were selected. 
Table~\ref{tab:test_matrices} shows the main features
of their coefficient matrices.
Note that all of them are rank-deficient. 
%
When the precision of some other methods on those problems 
was reported in other works, we included their results here.

\begin{table}[ht!]
\tfvspace
\begin{center}
\small
\begin{tabular}{lrrrr} \hline
  Matrix &
  \multicolumn{1}{c}{$m$} &
  \multicolumn{1}{c}{$n$} &
  \multicolumn{1}{c}{nnz} &
  \multicolumn{1}{c}{rank} \\ \hline \hline
  162bit       &  3,606 &   3,597 &    37,118 &  3,460 \\ 
  176bit       &  7,441 &   7,431 &    82,270 &  7,110 \\ 
  192bit       & 13,691 &  13,682 &   154,303 & 13,006 \\ 
  208bit       & 24,430 &  24,421 &   299,756 & 22,981 \\ 
  Maragal\_6   & 21,255 &  10,152 &   537,692 &  8,331 \\ 
  Maragal\_7   & 46,845 &  26,564 & 1,200,537 & 20,843 \\ \hline\hline
\end{tabular}
\end{center}
\bfvspace
\caption{Selected test matrices for the precision analysis.}
\label{tab:test_matrices}
\end{table}

In addition to the coefficient matrices $A$,
this matrix suite sometimes provides vectors $b$.
If the problem provides a right-hand side vector $b$, it is used. 
If not provided, three additional different types of vectors $b$ were generated. 
Next, we describe the four different scenarios assessed:

\begin{itemize}

\item
Scenario 1:
Vector $b$ is provided in the Suite Sparse Matrix Collection.

\item
Scenario 2:
Vector $b$ is set to ones.
In this case, the system $Ax=b$ does not have a known solution.

\item
Scenario 3:
Vector $b$ is generated as $b=Ax$, where $x$ is generated randomly.

\item
Scenario 4:
Vector $b$ is generated as $b=Ax$, where $x$ is generated randomly.
Additionally, $b$ is slightly perturbed.

\end{itemize}

\subsubsection{Scenario 1: Vector $b$ is provided}

Unfortunately, only two problems in the Suite Sparse Matrix Collection
included vector $b$: Maragal\_6 and Maragal\_7.
Table~\ref{tab:residual_b_provided}
shows the residuals obtained in these problems.
As can be easily seen,
the residuals are the same for all methods.

\begin{table}[ht!]
\tfvspace
\begin{center}
\small
\begin{tabular}{lrrrr} \hline
             & \multicolumn{4}{c}{Method} \\
  Matrix     & \multicolumn{1}{c}{\texttt{v33s}} &
               \multicolumn{1}{c}{\texttt{v34s}} &
               \multicolumn{1}{c}{\texttt{MKL\_GELSS}} & 
               \multicolumn{1}{c}{\texttt{MKL\_GELSY}} \\ \hline \hline
  Maragal\_6 & 4.53e-04 & 4.53e-04 & 4.53e-04  & 4.53e-04  \\ \hline
  Maragal\_7 & 3.71e-06 & 3.71e-06 & 3.71e-06  & 3.71e-06  \\ \hline \hline
\end{tabular}
\end{center}
\bfvspace
\caption{Residuals $\| A x - b \|_F$ when $b$ is provided.}
\label{tab:residual_b_provided}
\end{table}

Scott~\cite{scott2017using} solved these problems 
by using the iterative method LSMR 
based on a Cholesky-based preconditioner 
that modifies the diagonal of the unregularized matrix 
to obtain matrices that are easier to factorize. 
The residuals for the Maragal\_6 and Maragal\_7 problems 
obtained by that work are 1.069e+01 and 1.369e+01, respectively,
much higher than those obtained 
by both our implementations and the reference implementations.

Table~\ref{tab:norm_x_b_provided} shows the norm of the solutions $x$.
Recall that \texttt{v33s} performs the complete process,
whereas \texttt{v34s} skips the annihilation of $T_{12}$. 
As can be seen in the table, the norm in each case is the same for both variants, 
and nearly equal to the norm of the solution obtained by the MKL methods.

\begin{table}[ht!]
\tfvspace
\begin{center}
\small
\begin{tabular}{lrrrr}
  \hline
             & \multicolumn{4}{c}{Method} \\
  Matrix     & \multicolumn{1}{c}{\texttt{v33s}} & 
               \multicolumn{1}{c}{\texttt{v34s}} &
               \multicolumn{1}{c}{\texttt{MKL\_GELSS}} &
               \multicolumn{1}{c}{\texttt{MKL\_GELSY}} \\ \hline \hline
  Maragal\_6 &  1.00e+01 & 1.00e+01  & 9.83e+00  & 9.83e+00     \\ \hline
  Maragal\_7 &  1.96e+01 & 1.96e+01  & 1.81e+01  &  1.81e+01       \\ \hline
  \hline
\end{tabular}
\end{center}
\bfvspace
\caption{Norm of the solution $\| x \|_F$ when $b$ is provided.
}
\label{tab:norm_x_b_provided}
\end{table}

\subsubsection{Scenario 2: Vector $b$ is set to ones}

In this scenario, vector $b$ was set to ones. 
Hence, the system does not have a known solution. 
Table~\ref{tab:residual_b_ones} shows 
the residuals of the solution for every test matrix. 
As can be seen, the residuals of the four methods 
were the same with three digits of precision.

\begin{table}[ht!]
\tfvspace
\begin{center}
\small
\begin{tabular}{lrrrr}
  \hline
             & \multicolumn{4}{c}{Method} \\
  Matrix     & \multicolumn{1}{c}{\texttt{v33s}} &
               \multicolumn{1}{c}{\texttt{v34s}} &
               \multicolumn{1}{c}{\texttt{MKL\_GELSS}} & 
               \multicolumn{1}{c}{\texttt{MKL\_GELSY}} \\ \hline \hline
  162bit     & 6.22e-01 & 6.22e-01 & 6.22e-01  & 6.22e-01  \\ \hline
  176bit     & 8.04e-01 & 8.04e-01 & 8.04e-01  & 8.04e-01  \\ \hline
  192bit     & 1.28e+00 & 1.28e+00 & 1.28e+00  & 1.28e+00  \\ \hline
  208bit     & 1.62e+00 & 1.62e+00 & 1.62e+00  & 1.62e+00  \\ \hline
  Maragal\_6 & 9.39e+01 & 9.39e+01 & 9.39e+01  & 9.39e+01  \\ \hline
  Maragal\_7 & 1.33e+02 & 1.33e+02 & 1.33e+02  & 1.33e+02  \\ \hline
  \hline
\end{tabular}
\end{center}
\bfvspace
\caption{Residual $\| A x - b \|_F$ of the execution when $b$ is set to ones.}
\label{tab:residual_b_ones}
\end{table}

When compared to the results obtained 
by Cerdán~\cite{cerdan2020preconditioners}, 
where they also solve the system using a vector $b$ set to ones, 
it can be seen that the residuals are the same 
with their Update Preconditioned Method (UPD) and 
also the non-updated preconditioner method (M) 
based on the Incomplete Cholesky (IC) factorization (Table 2 of the reference~\cite{cerdan2020preconditioners}).

Table~\ref{tab:norm_x_b_ones} shows the norm of the solutions $x$.
The norms are all almost equal, with only small differences between our methods and the MKL methods in some cases.

\begin{table}[ht!]
\tfvspace
\begin{center}
\small
\begin{tabular}{lrrrr}
  \hline
             & \multicolumn{4}{c}{Method} \\
  Matrix     & \multicolumn{1}{c}{\texttt{v33s}} & 
               \multicolumn{1}{c}{\texttt{v34s}} &
               \multicolumn{1}{c}{\texttt{MKL\_GELSS}} &
               \multicolumn{1}{c}{\texttt{MKL\_GELSY}} \\ \hline \hline
  162bit     & 1.08e+01 & 1.08e+01 & 1.04e+01 & 1.04e+01  \\ \hline
  176bit     & 1.50e+01 & 1.50e+01 & 1.47e+01 & 1.47e+01  \\ \hline
  192bit     & 2.15e+01 & 2.15e+01 & 2.12e+01 & 2.12e+01  \\ \hline
  208bit     & 2.49e+01 & 2.49e+01 & 2.47e+01 & 2.47e+01 \\ \hline
  Maragal\_6 & 3.64e+05 & 3.64e+05 & 3.64e+05 & 3.64e+05 \\ \hline
  Maragal\_7 & 1.15e+06 & 1.15e+06 & 1.15e+06 & 1.15e+06\\ \hline
  \hline
\end{tabular}
\end{center}
\bfvspace
\caption{Norm of the solution $\| x \|_F$ when $b$ is set to ones.
}
\label{tab:norm_x_b_ones}
\end{table}

\subsubsection{Scenario 3: Systems with known solution}

In this scenario, 
vector $b$ has been computed as $b=Ax$,
where matrix $x$ is randomly generated with values between 0 and 1.
Table~\ref{tab:residual_x_known} shows the residuals obtained.
As can be seen, in this case, the residuals are much lower 
than in the previous scenario.
The \texttt{v33s} and \texttt{v34s} methods consistently get better residuals than 
those of the \texttt{MKL\_GELSS} method.
The \texttt{v33s} and \texttt{v34s} methods get similar residuals to 
those of the \texttt{MKL\_GELSY} method:
In some problems, they are slightly higher,
and in some problems, they are slightly lower (192bit and 208bit).

\begin{table}[ht!]
\tfvspace
\begin{center}
\small
\begin{tabular}{lrrrr}
\hline
             & \multicolumn{4}{c}{Method} \\
  Matrix     & \multicolumn{1}{c}{\texttt{v33s}} &
               \multicolumn{1}{c}{\texttt{v34s}} &
               \multicolumn{1}{c}{\texttt{MKL\_GELSS}} & 
               \multicolumn{1}{c}{\texttt{MKL\_GELSY}} \\ \hline \hline
162bit     & 1.66e-12 & 1.66e-12 & 6.02e-12  & 1.18e-12  \\ \hline
176bit     & 3.69e-12 & 3.69e-12 & 1.11e-11  & 3.42e-12  \\ \hline
192bit     & 6.02e-12 & 6.02e-12 & 1.72e-10  & 8.30e-12  \\ \hline
208bit     & 7.51e-12 & 7.51e-12 & 1.51e-10  & 1.19e-11  \\ \hline
Maragal\_6 & 1.55e-12 & 1.55e-12 & 1.83e-12  & 6.09e-13  \\ \hline
Maragal\_7 & 2.79e-12 & 2.79e-12 & 4.00e-12  & 1.44e-12  \\ \hline
\hline
\end{tabular}
\end{center}
\bfvspace
\caption{Residual $\| A x - b \|_F$ of the execution when $x$ is known.}
\label{tab:residual_x_known}
\end{table}

Table~\ref{tab:norm_x_known} shows the norm of the obtained solution $X$.
As shown, the norms are similar.
There is no difference between \texttt{v33} ($T_{12}$ is annihilated), and
\texttt{v34} ($T_{12}$ is not annihilated).

\begin{table}[ht!]
\tfvspace
\begin{center}
\small
\begin{tabular}{lrrrr}
\hline
           & \multicolumn{4}{c}{Method} \\
Matrix     & \multicolumn{1}{c}{\texttt{v33s}} & 
             \multicolumn{1}{c}{\texttt{v34s}} & 
             \multicolumn{1}{c}{\texttt{MKL\_GELSS}} & 
             \multicolumn{1}{c}{\texttt{MKL\_GELSY}} \\ \hline \hline
162bit     & 3.67e+01 & 3.67e+01 & 3.39e+01 & 3.39e+01       \\ \hline
176bit     & 5.20e+01 & 5.20e+01 & 4.85e+01 & 4.85e+01     \\ \hline
192bit     & 6.82e+01 & 6.82e+01 & 6.61e+01 & 6.61e+01     \\ \hline
208bit     & 8.95e+01 & 8.95e+01 & 8.76e+01 & 8.76e+01       \\ \hline
Maragal\_6 & 6.08e+01 & 6.08e+01 & 3.93e+01 & 3.93e+01      \\ \hline
Maragal\_7 & 8.36e+01 & 8.36e+01 & 5.84e+01 & 5.84e+01      \\ \hline
\hline
\end{tabular}
\end{center}
\bfvspace
\caption{Norm of the solution $\| x \|_F$ when $X$ is known.}
\label{tab:norm_x_known}
\end{table}

To conclude this scenario,
when the solution of the system is known, 
our method based on the randUTV provides 
consistently good residuals 
compared to other high-performance linear solvers 
that work on rank-deficient matrices stored in main memory.

\subsubsection{Scenario 4: Systems with vector $b$ perturbed}

In this scenario, 
vector $b$ has been computed as $b=Ax$,
where vector $x$ is randomly generated with values between 0 and 1.
In contrast with the previous case,
the right-hand side $b$ has been perturbed in the following way: 
$10\%$ of the matrix elements (in random positions) have been set to 
a $99.9\%$ of the original value. 

Table \ref{tab:residual_x_known_b_perturbed3} shows the results obtained. 
As shown, the four methods also obtain the exact same residuals.
Compared with other cases,
the residuals are higher than in Scenario 3 
since vector $b$ has been perturbed.
However, the residuals are lower than those obtained 
in Scenario 2 ($b$ set to ones).

\begin{table}[ht!]
\tfvspace
\begin{center}
\small
\begin{tabular}{lrrrr}
\hline
           & \multicolumn{4}{c}{Method} \\
Matrix     & \multicolumn{1}{c}{\texttt{v33s}} & 
             \multicolumn{1}{c}{\texttt{v34s}} & 
             \multicolumn{1}{c}{\texttt{MKL\_GELSS}} & 
             \multicolumn{1}{c}{\texttt{MKL\_GELSY}} \\ \hline \hline
162bit     & 1.94e-02 & 1.94e-02 & 1.94e-02 & 1.94e-02 \\ \hline
176bit     & 3.32e-02 & 3.32e-02 & 3.32e-02 & 3.32e-02 \\ \hline
192bit     & 4.95e-02 & 4.95e-02 & 4.95e-02 & 4.95e-02 \\ \hline
208bit     & 6.77e-02 & 6.77e-02 & 6.77e-02 & 6.77e-02 \\ \hline
Maragal\_6 & 1.73e-02 & 1.73e-02 & 1.73e-02 & 1.73e-02 \\ \hline
Maragal\_7 & 2.30e-02 & 2.30e-02 & 2.30e-02 & 2.30e-02  \\ \hline
\hline
\end{tabular}
\end{center}
\bfvspace
\caption{Residual $\| A x - b \|_F$ of the execution 
when $x$ is known and $b$ has been perturbed (0.1\%).}
\label{tab:residual_x_known_b_perturbed3}
\end{table}

Table \ref{tab:norm_x_known_b_perturbed} shows 
the norm of the solutions for this scenario. 
The results show a similar behaviour to all the previous scenarios, 
where the MKL methods obtain a slightly lower norm, but are not significant.

\begin{table}[ht!]
\tfvspace
\begin{center}
\small
\begin{tabular}{lrrrr}
\hline
           & \multicolumn{4}{c}{Method} \\
Matrix     & \multicolumn{1}{c}{\texttt{v33s}} & 
             \multicolumn{1}{c}{\texttt{v34s}} & 
             \multicolumn{1}{c}{\texttt{MKL\_GELSS}} & 
             \multicolumn{1}{c}{\texttt{MKL\_GELSY}} \\ \hline \hline
162bit     & 3.67e+01  & 3.67e+01  &  3.39e+01     &  3.39e+01     \\ \hline
176bit     & 5.20e+01  & 5.20e+01  &  4.85e+01     &  4.85e+01       \\ \hline
192bit     & 6.82e+01  & 6.82e+01  &  6.61e+01     &  6.61e+01      \\ \hline
208bit     & 8.95e+01  & 8.95e+01  &  8.76e+01     &  8.76e+01         \\ \hline
Maragal\_6 & 1.22e+02  & 1.22e+02  &  1.12e+02     &  1.12e+02      \\ \hline
Maragal\_7 & 4.18e+02  & 4.18e+02  &  4.13e+02     &  4.13e+02      \\ \hline
\hline
\end{tabular}
\end{center}
\bfvspace
\caption{Norm of the solution $\| x \|_F$ when $x$ is known and $b$ has been perturbed (0.1\%).
}
\label{tab:norm_x_known_b_perturbed}
\end{table}

To conclude this precision analysis,
we can claim that the numerical stability of the linear least-squares solvers 
based on the randUTV factorization is up to par
with state-of-the-art widely used methods from high-performance high-quality libraries
such as Intel MKL.
Moreover,
the precisions of our methods are also competitive 
with those of other methods reported in the scientific literature
\cite{scott2017using,cerdan2020preconditioners}. 
All these results allow us to claim that our methods as very suitable
options for rank-deficient systems,
especially if they are of large dimensions.

\subsection{Performance analysis}
\label{subsec:performance_analysis}

This subsection contains the performance assessment 
of the new codes described in previous sections,
as well as some state-of-the-art codes included and assessed as a reference.
To make this study more thorough,
the following three architectures have been assessed:

\begin{itemize}

\item
\texttt{volta1 CPU}:
A server with 40 cores (Intel Xeon Gold 6138 CPU) at 2.0 GHz
and a main memory of 97 GB.
The operating system is Debian GNU/Linux 10 (buster).
The C compiler is GNU gcc (Debian 8.3.0-6) 8.3.0.
The computational library is 
Intel oneAPI Math Kernel Library Version 2023.0-Product Build 20221128 
for Intel 64 architecture applications.

\item
\texttt{volta1 GPU}:
Previous machine with a Tesla V100 with 32 GiB of GPU memory.
In this case, the Tesla V100 is employed for computational tasks
instead of the Intel CPU cores.
The computational library is CUDA Release 11.7.

\item
\texttt{mb CPU}:
A server with 64 cores (Intel Xeon Platinum 8358 CPU at 2.6 GHz)
and a main memory of 263 GB.
Note that since it has a larger main memory than the previous one, 
to exercise the secondary storage system,
matrix sizes must be larger.
The operating system is Debian GNU/Linux 11 (bullseye).
The C compiler is GNU gcc (Debian 10.2.1-6) 10.2.1 20210110.
The computational library is 
Intel oneAPI Math Kernel Library Version 2023.0-Product Build 20221128 
for Intel 64 architecture applications.
This server has been chosen because it has three different disks:

  \begin{itemize}
  \item
  \textit{Spinning disk}: A traditional spinning disk.
  \item
  \textit{SSD disk}: A NVMe (Non-Volatile Memory express) disk.
  \item
  \textit{Optane disk}: A virtual disk implemented on a 1-GiB Intel Optane memory.
  \end{itemize}
  
\end{itemize}

In all the experiments below,
we have used $q=0$ (number of steps in the power iteration process),
since this number has been usually enough to reveal the rank.

When comparing several methods with different computational costs,
the gigaflop rates (gigaflop per second) 
only report about the efficiency of the codes,
and are not useful to assess the overall cost (total time).
Since we are going to compare very different methods (QR, SVD, and randUTV) 
for solving linear systems
with different computational costs, 
all the graphics below employ the scaled time in the vertical axis,
unless otherwise stated.
The scaled time is computed as 
the product of the overall time and $10^{12}$ 
(to make it more manageable) 
divided by $n^3$,
where $n$ is the number of columns,
since the computational cost of all methods 
being assessed in this work is $O(n^3)$.

In all the experiments the block size employed is 10\,240,
since this is usually the most performant one.
Accordingly,
all the matrix dimensions assessed in this section 
are a multiple of this block size.
Nevertheless, the methods and all the implementations work
on any block size and matrix dimension,
even if they are not multiple.

Unless stated otherwise,
in all the experiments when solving linear systems below,
dimension $k$ 
(the number of columns of matrix $B$ or 
the number of linear systems simultaneously being solved)
is 1\,024.

The block cache size in those variants that use it
is 20 GB.
Since the dimension of square blocks 
(those storing data from matrices $A$, $U$, and $V$)
is usually $10\,240 \times 10\,240$, 
the storage of just one of them requires 800 MB
when working with double precision.
In this case, the block cache could store up to 25 square blocks.
In case of blocks of $X$ and $B$, many more blocks can be stored
since their column dimensions are usually smaller.

Analogously,
unless stated otherwise,
when assessing rank-deficient matrices,
the matrix rank for a matrix of dimension $1\,024 p$ is $1\,000 p$.
That is,
the rank for a matrix of dimension 81\,920 is 80\,000,
the rank for a matrix of dimension 92\,160 is 90\,000, and so on.
That is, about 2.3 \% of the columns are linearly dependent.

The coefficient matrices being assessed in this performance subsection 
are built as follows.
When a matrix of dimension $m \times n$ and numerical rank $r$ must be built,
the first step is to build a row block of dimension $r \times n$
with numerical rank $r$.
This is achieved by generating a random block of those dimensions and 
then making it diagonal-dominant.
The second step is 
to replicate the first block scaled by a random factor
as many times as needed to fill the $m$ rows.
Note that though this method is very simple,
it can be easily scaled up to dimensions so large 
that the matrices do not fit in main memory.

\subsubsection{Assessment of the block cache systems}
\label{subsubsec:block_cache_systems}

\begin{table}[ht!]
\tfvspace
\begin{center}
\small
\begin{tabular}{lrrrrr} \hline & 
\multicolumn{1}{c}{\texttt{v23t}} & 
\multicolumn{1}{c}{\texttt{v23c}} & 
\multicolumn{1}{c}{\texttt{v23d}} & 
\multicolumn{1}{c}{\texttt{v23e}} &
\multicolumn{1}{c}{\texttt{v23x}}  \\ \hline \hline
\# Disk Reads        &  3\,799    &  2\,092   &  2\,072   &  1\,567   &  1\,688   \\ \hline
\# Disk Writes       &  2\,945    &  1\,576   &  1\,558   &  1\,259   &  1\,342   \\ \hline
\hline
\end{tabular}
\end{center}
\bfvspace
\caption{Number of disk operations
for several OOC variants on the multicore system \texttt{volta1 CPU}.}
\label{tab:cache_multicore}
\end{table}

Table~\ref{tab:cache_multicore} shows the number of disk operations
for several variants when processing 
a matrix of dimension $92\,160 \times 92\,160$ with rank $90\,000$
on \texttt{volta1 CPU}.
Recall that variant \texttt{v23t} does not employ any block cache,
variant \texttt{v23c} employs the basic cache with 
a LRU (Least-Recently-Used) policy,
variant \texttt{v23d} employs a more sophisticated cache with a LRU policy, 
and variant \texttt{v23e} employs a more sophisticated cache with a LFU 
(Latest-in-the-Future-to-be-Used) policy.
The last variant \texttt{v23x} employs the same LFU cache
plus overlapping of computation and disk I/O.

As can be seen, the use of a cache greatly reduces the number of 
disk reads (45 \%, from 3\,799 to 2\,092)
and disk writes (46 \%, from 2\,945 to 1\,576).
A more advanced cache system further reduces 
the previous number of disk reads (from 2\,092 to 2\,072) and 
disk writes (from 1\,576 to 1\,558).
The new cache system with the LFU eviction policy
reduces even more the number of disk reads (from 2\,072 to 1\,567)
and disk writes (from 1\,558 to 1\,259).

On the other side, although variant \texttt{v23x}, 
which employs overlapping of I/O and computation,
also uses the same sophisticated cache system as variant \texttt{v23e},
the number of disk operations performed slightly increases.
This deserves a close analysis.
When overlapping disk I/O and computation,
the I/O thread reads blocks in advance.
Obviously,
from the time a block is read by the I/O thread 
until that block is processed by the computational thread,
that block cannot be chosen to be discarded 
when trying to load a new block in the cache.
This makes that the cache usually contains a set of blocks 
pending to be ``used'' that cannot be discarded,
thus reducing the effective size of the cache.
If the size of the cache is reduced, 
the effectiveness of the cache is also reduced,
thus slightly increasing the number of disk reads and writes.

In overall,
the reduction of the number of disk reads and disk writes
achieved by the variant with overlapping
with respect to the non-cache variant
is 56 \% (from 3\,799 to 1\,688) and 
54 \% (from 2\,945 to 1\,342). 

Identical results were obtained for the variants 
developed for the GPU architecture,
with only slight differences 
in the variant with overlapping 
because of the indeterminism introduced by the concurrency of 
the computational thread and the disk I/O thread.

\subsubsection{Decomposed times of variants}
\label{subsubsec:decomposed_times}

\begin{table}[ht!]
\tfvspace
\begin{center}
\small
\begin{tabular}{lrrrrrr} \hline & 
\multicolumn{1}{c}{\texttt{v21t}} & 
\multicolumn{1}{c}{\texttt{v22t}} & 
\multicolumn{1}{c}{\texttt{v23t}} & 
\multicolumn{1}{c}{\texttt{v23s}} & 
\multicolumn{1}{c}{\texttt{v33t}} & 
\multicolumn{1}{c}{\texttt{v33s}} \\ \hline \hline
Disk I/O total 
time & 5\,158.6  & 5\,685.4  & 3\,984.2  & 2\,005.3 & 3\,996.8 & 1\,957.6 \\ \hline
GPU I/O 
time & ---     & ---     & ---     & ---    & 1\,255.9 & 733.2  \\ \hline
Computational 
time & 9\,991.3  & 8\,425.1  & 6\,701.8  & 6\,559.0 & 2\,442.5 & 3\,853.6 \\ \hline
Added 
time & 15\,149.9 & 14\,110.5 & 10\,685.9 & 8\,564.3 & 7\,695.2 & 6\,544.5 \\ \hline
Real 
time & 15\,389.0 & 14\,366.0 & 10\,862.0 & 6\,639.9 & 7\,799.2 & 4\,906.0 \\ \hline
\hline
\end{tabular}
\end{center}
\bfvspace
\caption{Decomposed and real times in seconds for several OOC variants.}
\label{tab:decomposed_times}
\end{table}

Table~\ref{tab:decomposed_times}
compares the decomposed and real times of several of 
our least-squares variants 
when solving a linear system of dimension $92\,160 \times 92\,160$.
Recall that variants \texttt{v2xx} are CPU-based, whereas
variants \texttt{v3xx} are GPU-based.
The first data row shows the time spent in disk I/O
(transferring data between disk and main memory).
The second data row shows the time spent in GPU I/O
(transferring data between main memory and the GPU memory),
which does not apply for CPU variants.
The third data row shows the time spent in computations
(either CPU-based or GPU-based).
The fourth data row shows the addition of the previous three times.
The fifth data row shows the real time.

As can be seen, the real time drops 
from 15\,389.0 seconds (traditional variant \texttt{v21t}) 
to 6\,639.9 (fastest CPU variant \texttt{v23s})
to 4\,906.0 (fastest GPU variant \texttt{v33s}).
Concretely,
by tuning the basic computational tasks,
the real time drops 7 \% 
(from 15\,389.0 in \texttt{v21t} to 14\,366.0 in \texttt{v22t}).
By not building matrix $U$ explicitly,
the real time drops 32 \% 
(from 14\,366.0 in \texttt{v22t} to 10\,862.0 in \texttt{v23t}).
By overlapping I/O and computation,
the real time drops 64 \% 
(from 10\,862.0 in \texttt{v22t} to 6\,639.9 in \texttt{v23t}).
The GPU variants further reduces the real time 
by reducing the computational cost.
In this case, the reduction is 35 \%
(from to 6\,639.9 to 4\,906.0).

\subsubsection{Comparison of elementary tasks}
\label{subsubsec:comparison_elementary_tasks}

\begin{table}[ht!]
\tfvspace
\begin{center}
\footnotesize
\begin{tabular}{lrrrrrr} \hline & 
\multicolumn{1}{c}{v21t} & 
\multicolumn{1}{c}{v22t} & 
\multicolumn{1}{c}{v23t} & 
\multicolumn{1}{c}{v23s} & 
\multicolumn{1}{c}{v33t} & 
\multicolumn{1}{c}{v33s} \\ \hline  \hline 
Comp\_De      &   2.346    &   1.885    &   1.926    &   1.759    &   0.752     &   1.193    \\
              &         16 &         16 &         16 &         16 &         16  &         16 \\ \hline 
Comp\_TD      &  13.750    &   4.946    &   5.005    &   5.247    &   0.987     &   1.256    \\
              &         72 &         72 &         72 &         72 &         72  &         72 \\ \hline 
Appl\_l\_De   &   2.160    &   2.393    &   2.020    &   1.673    &   0.366     &   0.840    \\
              &         36 &         36 &         44 &         44 &         44  &         44 \\ \hline 
Appl\_r\_De   &   2.734    &   2.331    &   2.249    &   2.089    &   0.413     &   0.718    \\
              &        216 &        216 &        144 &        144 &        144  &        144 \\ \hline 
Appl\_l\_TD   &   4.613    &   4.748    &   4.053    &   3.459    &   0.699     &   1.876    \\
              &        204 &        204 &        240 &        240 &        240  &        240 \\ \hline 
Appl\_r\_TD   &   5.671    &   4.767    &   4.959    &   4.098    &   0.760     &   1.715    \\
              &        972 &        972 &        648 &        648 &        648  &        648 \\ \hline 
Gemm\_nn      &   0.317    &   0.339    &   0.300    &   0.503    &   0.030     &   0.030    \\
              &        117 &        117 &        117 &        117 &        117  &        117 \\ \hline 
Gemm\_tn      &   1.164    &   1.126    &   1.375    &   1.281    &   0.331     &   0.557    \\
              &        365 &        365 &        284 &        284 &        284  &        284 \\ \hline 
Gemm\_aabt    &   1.496    &   1.476    &   1.487    &   1.672    &   0.370     &   0.500    \\
              &        117 &        117 &        117 &        117 &        117  &        117 \\ \hline 
Gemm\_abta    &   1.441    &   1.434    &   1.198    &   1.427    &   0.276     &   0.278    \\
              &         36 &         36 &         45 &         45 &         45  &         45 \\ \hline 
Gemm\_aab     &   1.510    &   1.474    &   --       &   --       &   --        &   --       \\
              &         81 &         81 &         -- &         -- &         --  &         -- \\ \hline 
Svd           &  92.793    &  93.890    &  94.976    & 109.462    & 151.100     & 194.047    \\
              &          9 &          9 &          9 &          9 &          9  &          9 \\ \hline 
Trsm\_lunn    &   0.163    &   0.136    &   0.144    &   0.193    &   0.022     &   0.022    \\
              &          9 &          9 &          9 &          9 &          9  &          9 \\ \hline 
Comp\_RZ      &   4.582    &   4.759    &   4.863    &   7.589    &   1.065     &   0.748    \\
              &          9 &          9 &          9 &          9 &          9  &          9 \\ \hline 
Appl\_r\_RZ   &   1.093    &   1.184    &   1.216    &   4.667    &   0.718     &   0.401    \\
              &        117 &        117 &        117 &        117 &        117  &        117 \\ \hline 
Normal        &   0.134    &   0.116    &   0.117    &   0.092    &   0.126     &   0.098    \\
              &         44 &         44 &         44 &         44 &         44  &         44 \\ \hline 
Keep\_upp     &   0.059    &   0.067    &   0.057    &   0.067    &   0.092     &   0.061    \\
              &          8 &          8 &          8 &          8 &          8  &          8 \\ \hline 
Zero          &   0.508    &   0.532    &   0.283    &   0.105    &   0.292     &   0.090    \\
              &         36 &         36 &         36 &         45 &         36  &         45 \\ \hline 
Disk\_Read    &   0.752    &   0.860    &   0.731    &   0.704    &   0.732     &   0.826    \\
              &       4745 &       4745 &       3799 &       1687 &       3799  &       1666 \\ \hline 
Disk\_Write   &   0.426    &   0.430    &   0.409    &   0.608    &   0.412     &   0.438    \\
              &       3738 &       3738 &       2945 &       1344 &       2945  &       1328 \\ \hline 
GPU\_Read     & --         & --         & --         & --         &   0.172     &   0.053    \\
              & --         & --         & --         & --         &       3072  &       6197 \\ \hline 
GPU\_Write    & --         & --         & --         & --         &   0.154     &   0.058    \\ 
              & --         & --         & --         & --         &       4613  &       6431 \\ \hline 
\# of ops.:   & 10947      & 10947      & 8703       & 4999       & 25650       & 30452      \\ \hline
\hline 
\end{tabular}
\end{center}
\bfvspace
\caption{Average time in seconds (upper line) and 
total number of operations (lower line)
for every elementary task in several randUTV OOC variants.}
\label{tab:elementary_tasks}
\end{table}

Table~\ref{tab:elementary_tasks}
shows the average time in seconds (upper line) and 
the number of operations executed (lower line)
for every elementary task 
in several least-squares solver variants.
As said before,
variant \texttt{v21t}
is a usual and traditional high-performance implementations 
based on BLAS-3.
In contrast,
the rest of CPU variants
employ broader (more complex) computational kernels from the Intel MKL library
(such as \texttt{dlarfb}, \texttt{dgeqrf}, \texttt{dormqr}, etc.)
since they have greatly optimized by this company.
As can be seen,
the rest of CPU variants
achieve higher performances in orthogonal-related tasks such as 
\texttt{Comp\_De},
\texttt{Comp\_TD},
\texttt{Appl\_l\_De},
\texttt{Appl\_r\_De},
\texttt{Appl\_l\_TD},
\texttt{Appl\_r\_TD}, etc.
In some cases, the performance gain 
ranges from a small percentage (such as 16 \%)
to several times (such as 2.7 times).

Note that
the use of overlapping of I/O and computations
can slightly increase the computational task in some cases.
This slight increase is caused by some interaction 
between the data transfer and the computation.
Anyway, this increase is clearly compensated 
by the ``hiding'' of all the I/O cost.

\subsubsection{Comparison of variants}
\label{subsubsec:comparison_variants}

\begin{figure}[ht!]
\tfvspace
\begin{center}
\begin{tabular}{cc}
\includegraphics[width=0.45\textwidth]{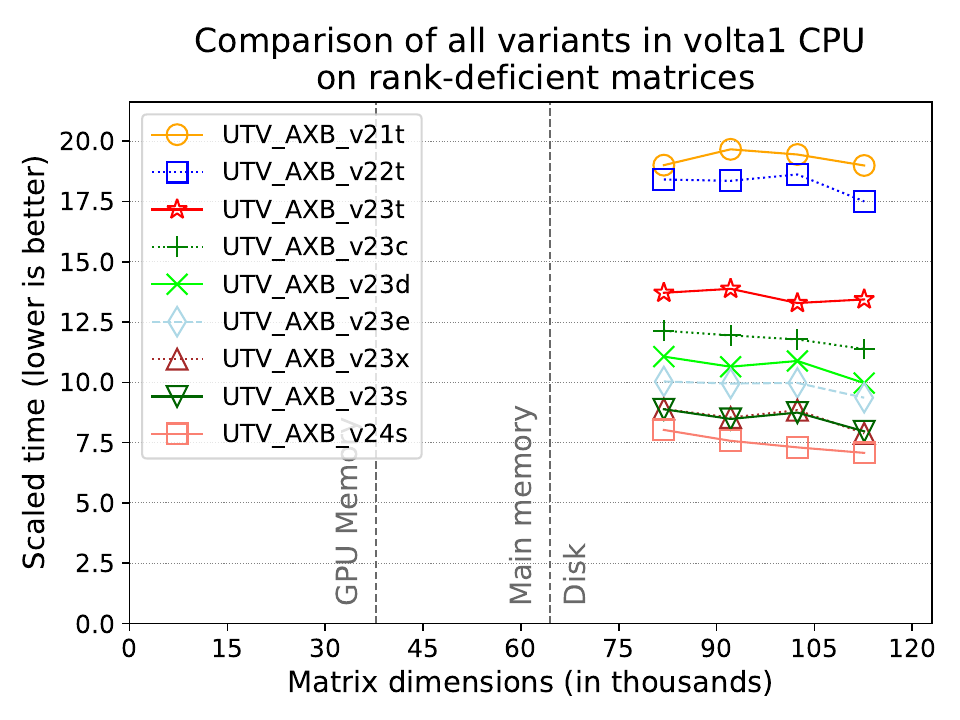} &
\includegraphics[width=0.45\textwidth]{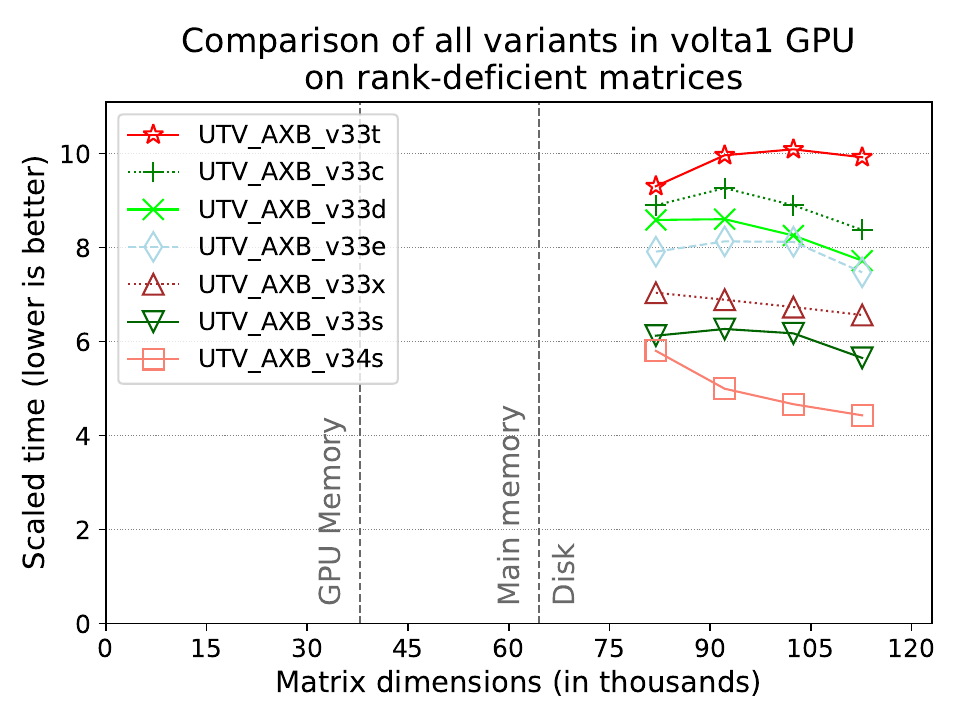} \\
\includegraphics[width=0.45\textwidth]{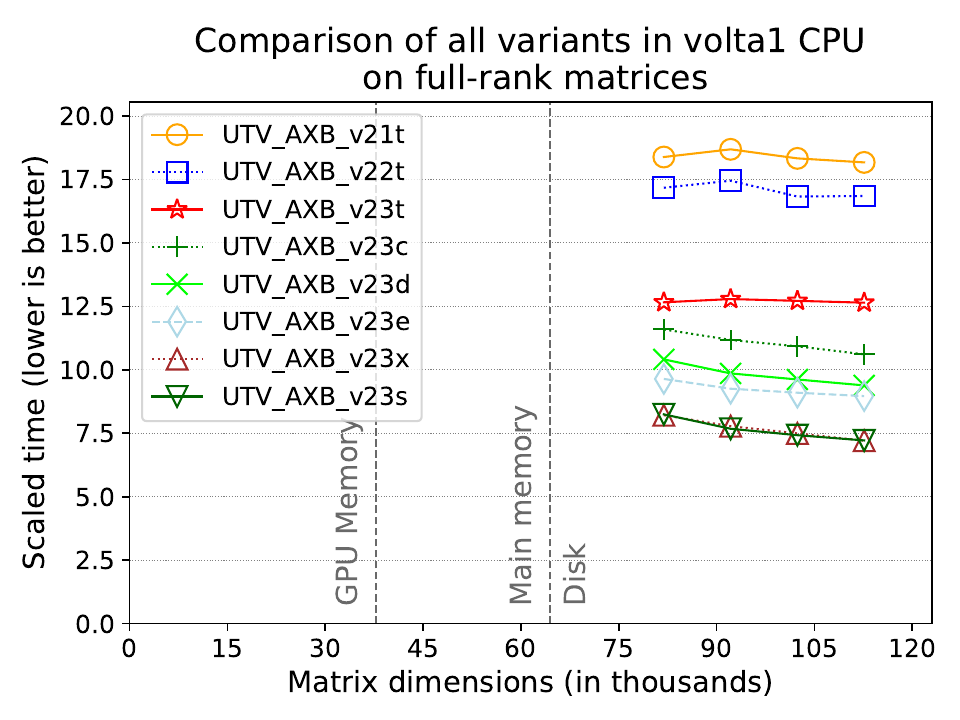} &
\includegraphics[width=0.45\textwidth]{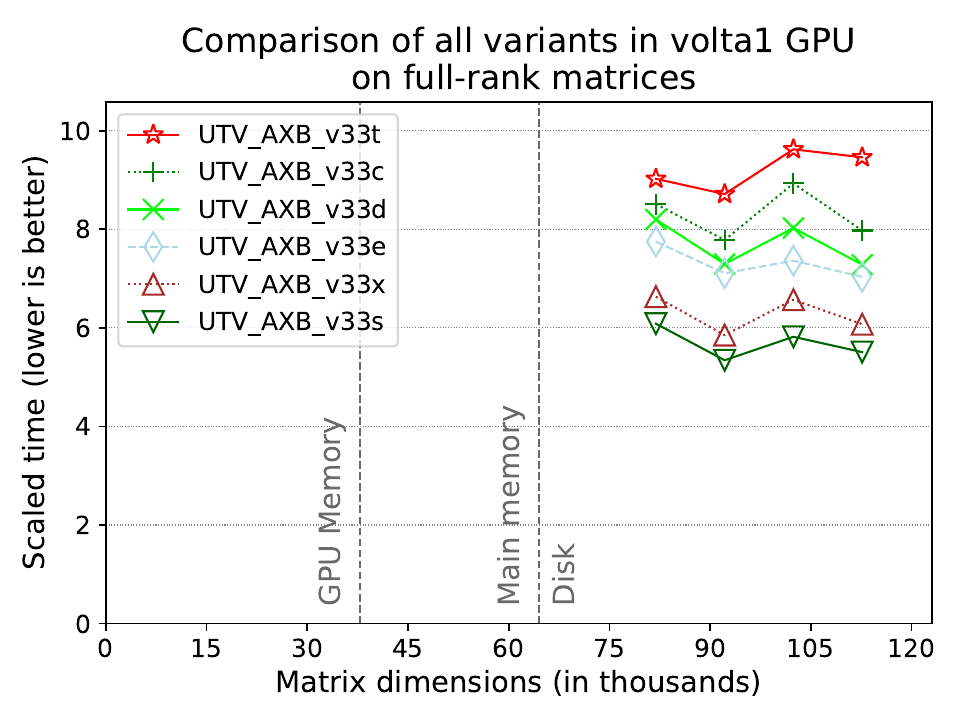} \\
\end{tabular}
\end{center}
\bfvspace
\caption{Scaled time versus matrix dimension 
for all the best least-squares solver variants.
The left plots show performances when performing computations on CPU, whereas
the right plots show performances when performing computations on GPU.
The top plots show results for rank-deficient matrices,  whereas
the bottom plots show results for full-rank matrices.}
\label{fig:comparison_variants}
\end{figure}

Figure~\ref{fig:comparison_variants} 
shows the scaled time of all of the variants enumerated above
with respect to the matrix dimensions.
This figure contains four plots as a $2 \times 2$ grid:
The top plots show the performances on rank-deficient matrices, whereas
the bottom plots show the performances on full-rank matrices.
The left plots show the performances on a multicore architecture
(\texttt{volta CPU}), whereas
the right plots show the performances on a GPU architecture
(\texttt{volta GPU}).
Obviously,
variants that annihilate $T_{12}$ are only shown for rank-deficient matrices,
since their performances are the same as the other variants 
if the matrix is full rank (and therefore $T_{12}$ is empty).

Note that each variant clearly outperforms the previous one.
The use of tuned computational CPU kernels in \texttt{v22t} variant
clearly increases performances with respect to \texttt{v21t} variant.
In this case, the speed gain derives from the computations exclusively.
The \texttt{v23t} variant
clearly outperform \texttt{v22t} variant
since the former applies the $U$ orthogonal matrix to matrix $B$ on-the-fly 
without explicitly building it, 
thus reducing the overall traffic and computation.
The cache variants 
\texttt{v23c} and \texttt{v33c}
achieve better performances in some cases
by reducing the amount of traffic between disk and main memory.
On the other hand,
variants 
\texttt{v23d} and \texttt{v33d} with the new optimized LRU cache
improve performances in all cases.
Finally, 
variants with the new LFU cache and overlapping of I/O and computations
(\texttt{v23x} and \texttt{v33x}) are even better.
Results of the \texttt{v23s} and \texttt{v33s} variants
show that the use of page-locked memory (\textit{pinned} memory) 
do not increase performances when working on CPU 
and slightly increases performances when working on GPU.
Variants \texttt{v24s} and \texttt{v34s} again increase performances,
in this case by saving the cost of annihilating block $T_{12}$.

The plots on GPU performances contain fewer variants 
since some lessons were taken advantage of from the CPU variants 
and the running times of the experiments are very large.

To conclude on these four plots,
a remarkable performance gain can be seen 
between the slowest variants and the fastest variants.
The performance gain is larger than two
when comparing the CPU best variant and the CPU worst variant.
On the other hand, the performance gain is about two
when comparing the GPU best variant and the GPU worst variant.

\subsubsection{Performances with respect to storage devices (disks)}
\label{subsubsec:comparison_disk}

\begin{figure}[ht!]
\tfvspace
\begin{center}
\begin{tabular}{cc}
\includegraphics[width=0.45\textwidth]{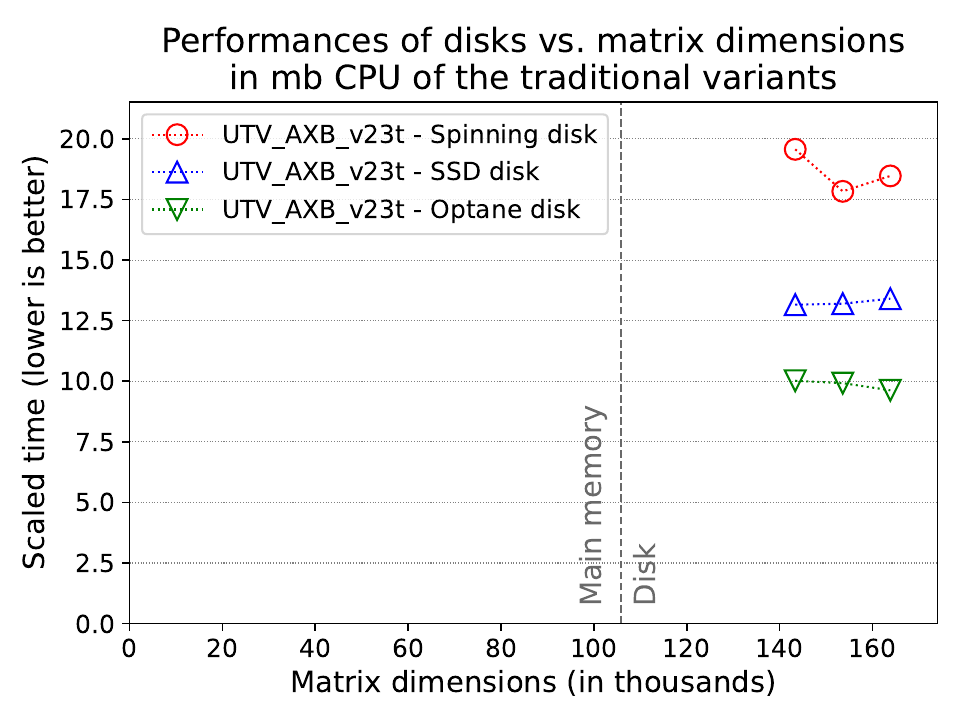}
&
\includegraphics[width=0.45\textwidth]{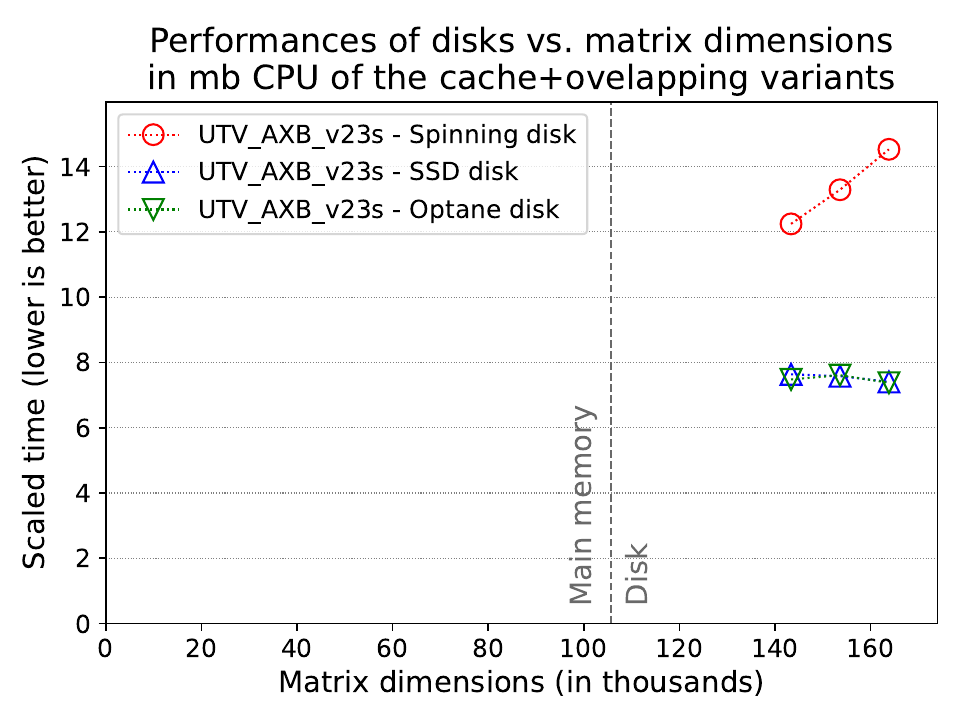}
\end{tabular}
\end{center}
\bfvspace
\caption{Scaled time versus matrix dimension
for the best implementations on several disks for rank-deficient matrices.
The left plot shows the results for the traditional variant, whereas
the right plot shows the results for the variant with cache and overlapping.
}
\label{fig:comparison_disk}
\end{figure}

Figure~\ref{fig:comparison_disk} shows the scaled time
with respect to matrix dimensions (size of the data)
for three different storage devices in \texttt{mb CPU}.
The left plot shows the results for the traditional variant
\texttt{v23t} on the three disks, whereas
the right plot shows the results for the variant \texttt{v23s} 
(cache and overlapping) on the three disks.
When working with the traditional variant (left plot),
the SSD disk greatly increases performances 
with respect to the spinning disk,
and again the Optane disk increases performances
with respect to the SSD disk.
On the other hand,
when working with the variant with cache and overlapping (right plot),
the SSD disk greatly increases performances 
with respect to the spinning disk.
In this case,
the Optane disk does not increase performances and 
obtains about the same performances than the SSD disk.

\subsubsection{Comparison between CPU and GPU implementations}
\label{subsubsec:comparison_cpu_gpu}

\begin{figure}[ht!]
\tfvspace
\begin{center}
\begin{tabular}{cc}
\includegraphics[width=0.45\textwidth]{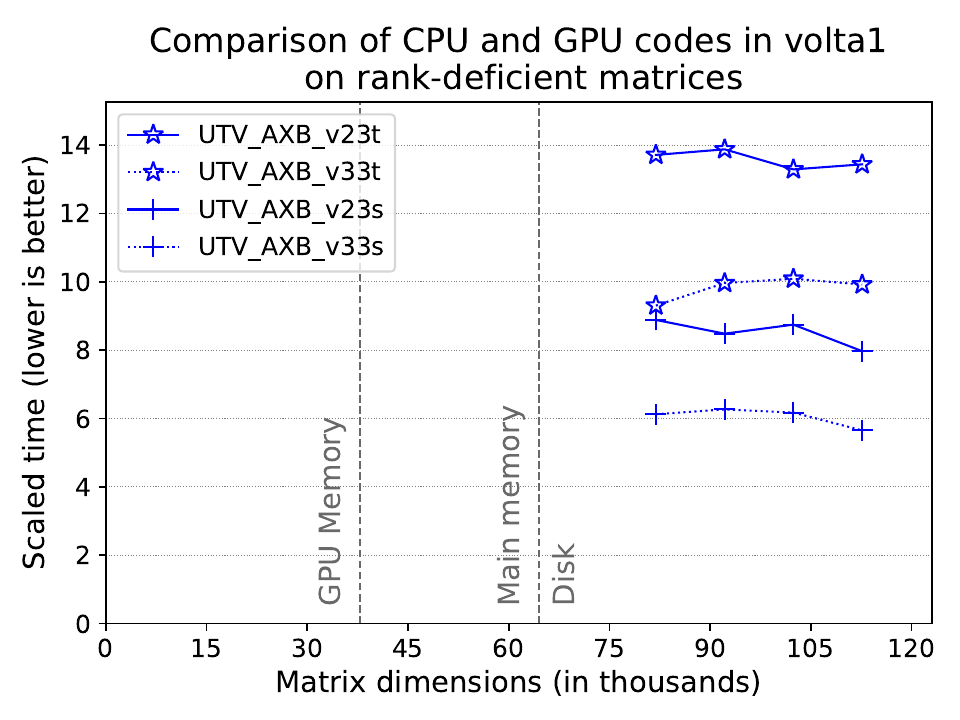} &
\includegraphics[width=0.45\textwidth]{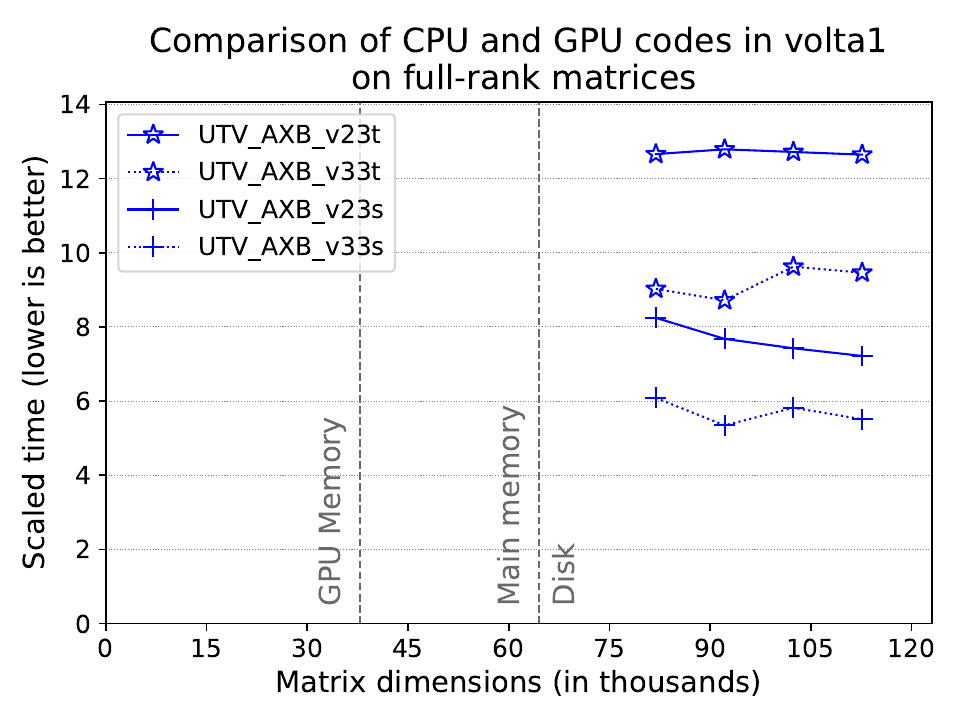} \\
\end{tabular}
\end{center}
\bfvspace
\caption{Scaled time versus matrix dimension for the best implementations 
based on CPU and GPU.
The left plot shows performances obtained 
when processing rank-deficient matrices, whereas
the right plot shows performances obtained 
when processing full-rank matrices.
}
\label{fig:comparison_cpu_gpu}
\end{figure}

Figure~\ref{fig:comparison_cpu_gpu}
compares the scaled times of 
the most-representative CPU-based implementations and
the most-representative GPU-based implementations
with respect to matrix dimensions.
The left plot shows results for rank-deficient matrices, whereas
the right plot shows results for full-rank matrices.
Comparing the traditional implementations based on the randUTV factorization,
there is a clear performance gap between 
the GPU implementation (\texttt{v33t})
and the CPU implementation (\texttt{v23t}).
Analogously, 
comparing the implementations with cache and overlapping 
based on the randUTV factorization,
there is a clear performance gap between 
the GPU implementation (\texttt{v33s})
and the CPU implementation (\texttt{v23s}).

\subsubsection{Performances with respect to rank}
\label{subsubsec:comparison_rank}

\begin{figure}[ht!]
\tfvspace
\begin{center}
\begin{tabular}{c}
\includegraphics[width=0.60\textwidth]{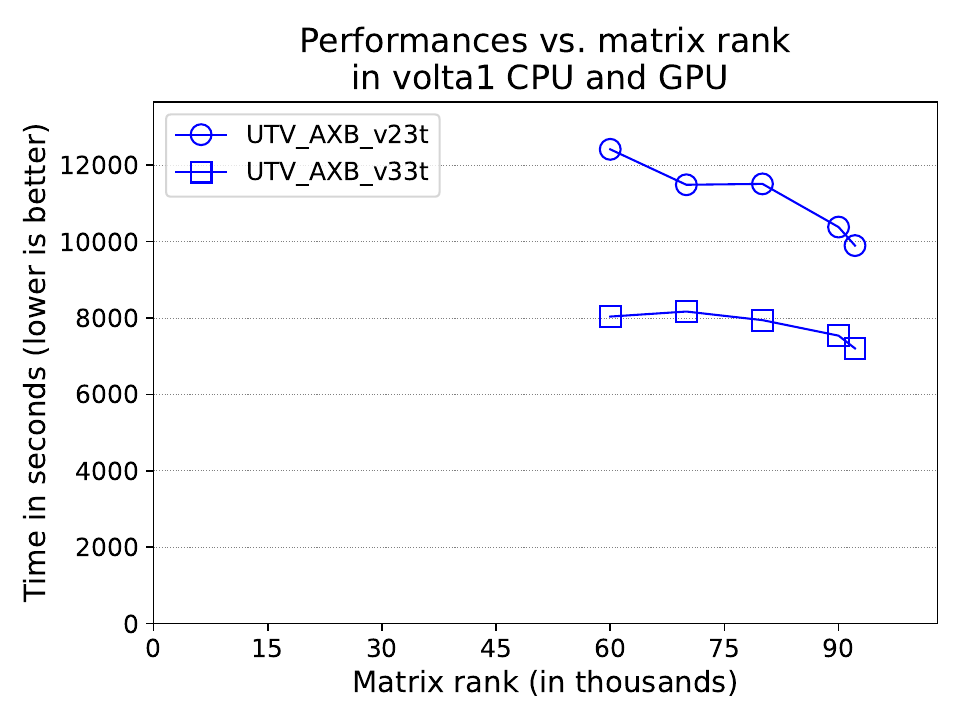}
\end{tabular}
\end{center}
\bfvspace
\caption{Times in seconds versus matrix rank 
for several implementations on matrices with different ranks.
}
\label{fig:comparison_rank}
\end{figure}

Figure~\ref{fig:comparison_rank}
shows the total times in seconds of several implementations
with respect to matrix rank.
Note that the times in this figure are absolute (in seconds) and not scaled.
Obviously, the total time depends on the matrix rank
since the smaller the rank is, 
the more elements in $T_{12}$ must be nullified,
the higher the computational cost is, and
therefore the larger the total time is.
In this case,
the GPU implementations depend slightly on the matrix rank.
On the other side,
the CPU implementations do clearly depend on the matrix rank.
With respect to the full-rank cost,
when the rank is close to half the matrix size,
the overall cost only grows about 25 \% on CPU 
and a much smaller percentage on GPU.

\subsubsection{Performances with respect to \textit{k}}
\label{subsubsec:comparison_k}

\begin{figure}[ht!]
\tfvspace
\begin{center}
\begin{tabular}{cc}
\includegraphics[width=0.45\textwidth]{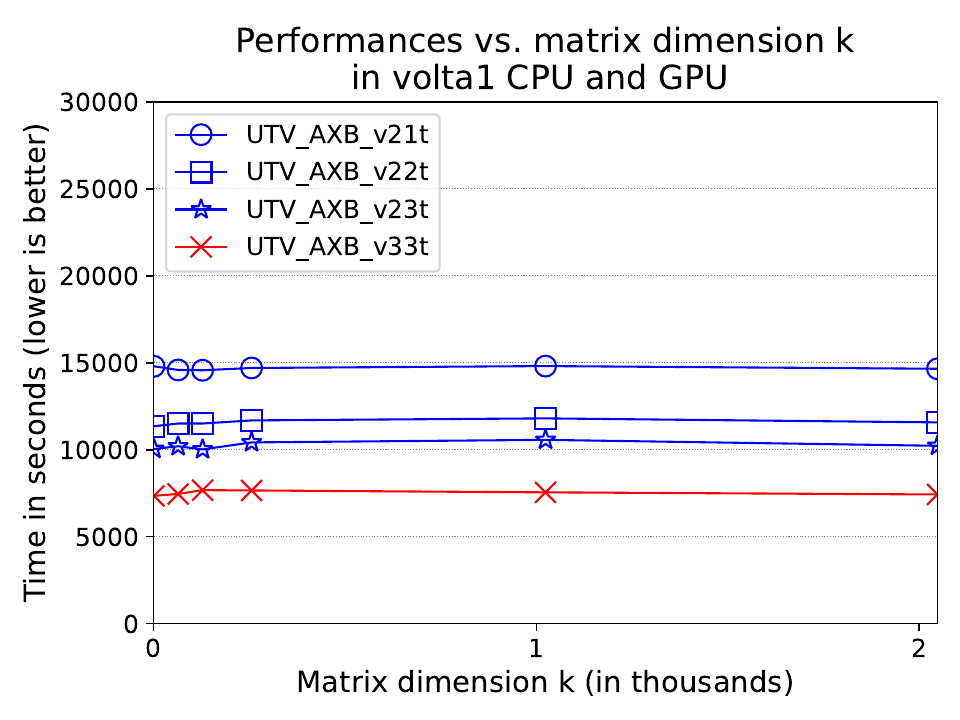}
&
\includegraphics[width=0.45\textwidth]{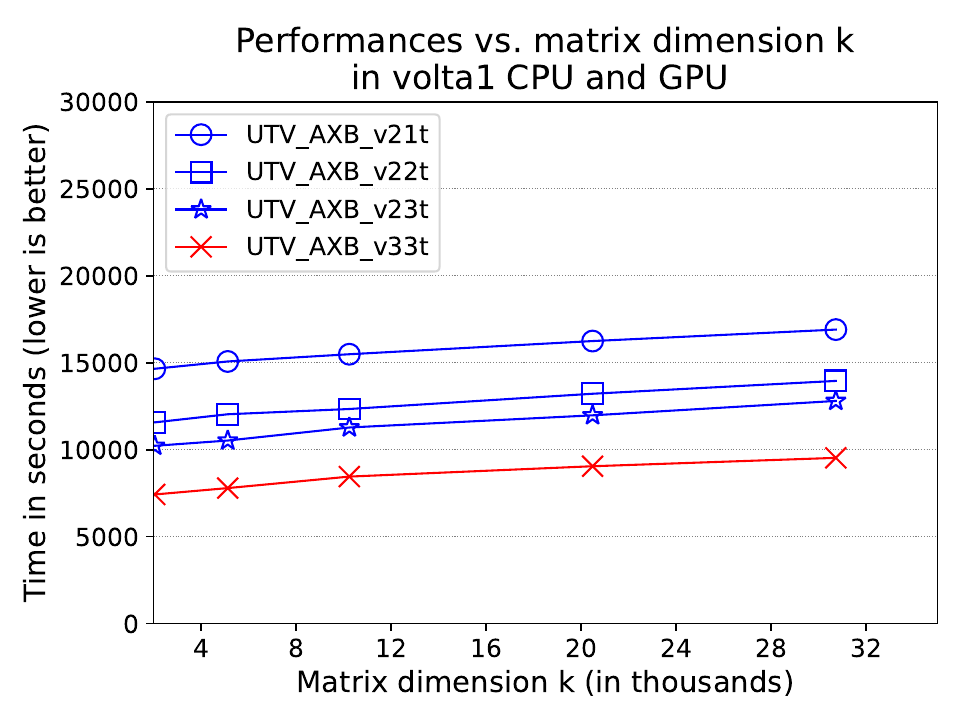} \\
\end{tabular}
\end{center}
\bfvspace
\caption{Time in seconds versus matrix dimension $k$
(number of columns in $B$)
for the best implementations for rank-deficient matrices.
The left plot shows data up to $k=2\,048$, whereas
the right plot shows data starting with $k=2\,048$.
}
\label{fig:comparison_k}
\end{figure}

When solving the problem $\min_X || A X - B ||$,
the number of columns of $B$ also affects the computational cost.
Figure~\ref{fig:comparison_k} 
shows the performances of the best new implementations 
with respect to the the number of columns in $B$ (called $k$).
In this case, note that this figure shows the total time in seconds
instead of the scaled time since matrix dimensions $m$ and $n$ do not vary.
Two plots have been employed to show the performances in more detail.
The left plot shows the times in seconds for small values of $k$ 
(up to 2\,048), 
and the right plot shows the times in seconds for larger values of $k$
(starting with 2\,048).
As shown,
the total time is about flat for up to 2\,048 columns in $B$,
and it only increases slightly starting with 2\,048 columns.
Note that the case with 30\,720 columns is only slightly slower
than the case with one column.
Examining the data in more detail,
the worst variant, \texttt{v21t},
requires 14\,796 and 16\,907 seconds 
for $k=1$ and $k=30\,720$, respectively,
which is a 14.3 \% increment when solving 30\,720 times as many systems.

\subsubsection{Comparison of In-Core codes and OOC codes}
\label{subsubsec:comparison_incore_ooc}

Several high-performance In-Core implementations
have been assessed and included here 
as a reference to our new Out-Of-Core implementations.
When high-performance In-Core implementations from commercial or 
open-source libraries were available, they were assessed.
When not available, we implemented them 
using high-performance commercial or open-source libraries.
We have assessed the following In-Core codes:

\begin{itemize}

\item
\texttt{MKL\_GELS}:
High-performance linear least squares solver \texttt{dGELS} from the Intel MKL library.
As it is based on the QR factorization, 
it does not work on rank-deficient matrices.
Obviously, this code has been only assessed with full-rank matrices.

\item
\texttt{MKL\_GELSY}:
High-performance linear least squares solver \texttt{dGELSY} from the Intel MKL library.
It can process rank-deficient matrices
since it is based on the column-pivoting QR factorization.

\item
\texttt{MKL\_GELSS}:
High-performance linear least squares solver \texttt{dGELSS} from the Intel MKL library.
It can process rank-deficient matrices
since it is based on the SVD factorization.

\item
\texttt{UTV\_AXB\_v11}:
High-performance implementation 
for solving systems with data stored in main memory (In-Core computations) 
based on the randUTV factorization.
It employs BLAS-3 from the Intel MKL library.
It can work on rank-deficient matrices.

\item
\texttt{CUSOLVER\_QR\_LS}:
High-performance implementation for GPU 
analogous to \texttt{MKL\_GELS}.
Since there is not a linear least-squares implementation 
in the NVIDIA cuSOLVER library
based on the QR factorization,
we have implemented it using the
\texttt{dgeqrf}, 
\texttt{dormqr}, and
\texttt{dtrsm} routines
from the high-performance NVIDIA cuSOLVER library.
As it is based on the QR factorization, 
it does not work on rank-deficient matrices.
In this case, this code has been assessed with full-rank matrices 
of the corresponding dimensions.

\item
\texttt{CUSOLVER\_SVD\_LS}:
High-performance implementation for GPU 
analogous to \texttt{MKL\_GELSS}.
This implementation is based on the high-performance NVIDIA cuSOLVER library.
Since there is not a linear least-squares implementation in cuSOLVER
based on the SVD factorization,
we have also implemented it using the
\texttt{xgesvd}, 
\texttt{dgemm}, and
\texttt{dtrsm} routines
from the high-performance NVIDIA cuSOLVER library.
As it is based on the SVD factorization, 
it can work on rank-deficient matrices.

\end{itemize}

\begin{figure}[ht!]
\tfvspace
\begin{center}
\begin{tabular}{cc}
\includegraphics[width=0.45\textwidth]{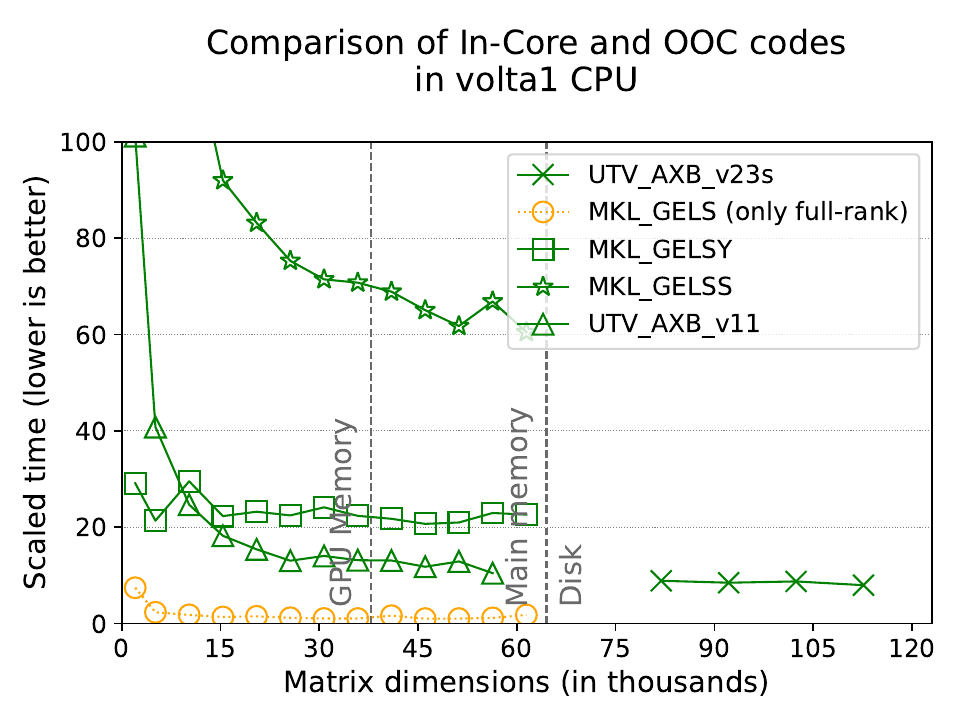}
&
\includegraphics[width=0.45\textwidth]{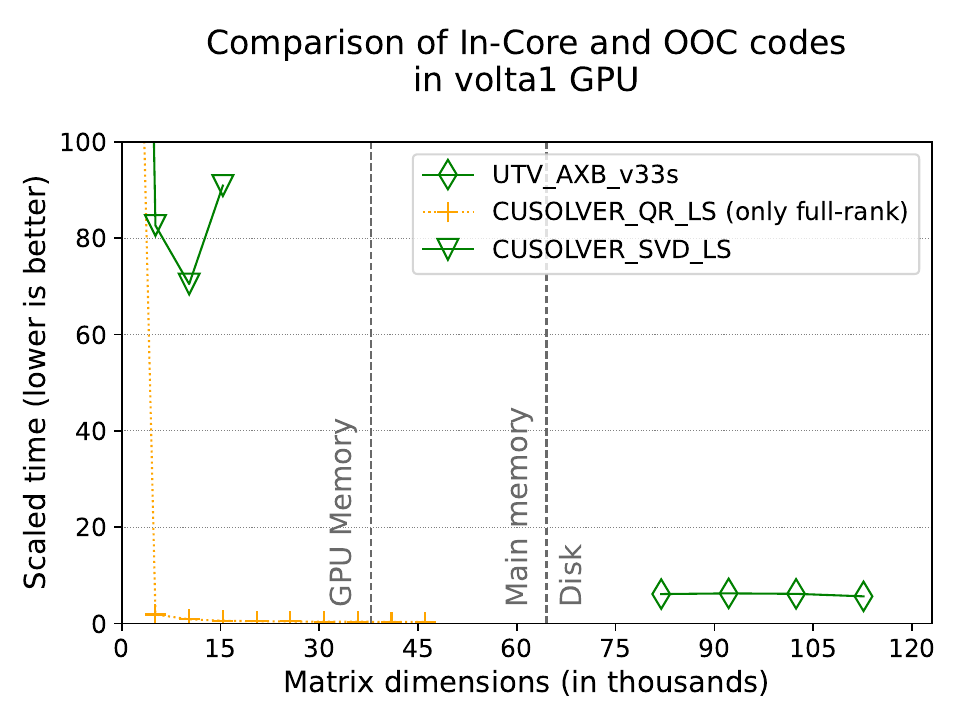}
\\
\end{tabular}
\end{center}
\bfvspace
\caption{Scaled time versus matrix dimensions
for the best implementations for rank-deficient matrices.
The left plot shows the performances of CPU codes, whereas 
the right plot shows the performances of GPU codes.
Both plots include both In-Core codes (left side of both plots) and 
Out-Of-Core codes (right side of both plots).
}
\label{fig:comparison_incore_ooc}
\end{figure}

Figure~\ref{fig:comparison_incore_ooc}
compares our new OOC implementations 
that can process data so large that must be stored in the hard drive 
versus state-of-the-art high-performance implementations 
that only work on data stored in main memory.
The left plot shows the performances on a CPU architecture, whereas
the right plot shows the performances on a GPU architecture.
Continuous lines are used for codes that can process rank-deficient matrices.
Non-continuous lines are used for codes than can only process full-rank matrices,
and are included just as a reference (and are usually faster). 
Codes with continuous lines have been assessed on rank-deficient matrices
with the corresponding ranks for the corresponding dimensions
($10\,000k$ rank for $10\,240k$ dimensions), as described before.
Codes with non-continuous lines have been assessed on full-rank matrices
with the corresponding dimensions
since they do not work on rank-deficient matrices.

In each plot, 
the lines in the left are the In-Core implementations,
that is, implementations 
that can only work on data stored 
in the more expensive (and limited) main memory.
Accordingly,
in each plot, 
the lines in the right are our new OOC implementations, 
which can work on matrices of any size, 
even though they do not fit in main memory and 
must be stored in the hard drive.

As shown,
our new OOC implementation on a CPU architecture 
are so fast that they can outperform high-performance implementations
from the Intel MKL library.
Our OOC implementation obtains a speed very similar to the In-Core
implementation that has all the data in main memory.
As can be seen on the right plot (GPU-based architectures),
our code is faster than an In-Core implementation based 
on the cuSOLVER library.
The right plot shows results 
for the implementation based on the SVD factorization from cuSOLVER
up to $n=15k$ 
since its convergence fails for $n \geq 20k$.
Note that in both plots, 
implementations based on the QR factorization
are obviously much faster, 
but you must recall that they only work on full-rank matrices.

\section{Conclusions}
\label{sec:conclusions}

Solving very large linear systems of equations
is a crucial task in many problems from science and technology.
In many cases, 
the coefficient matrix of the system is rank-deficient,
and usual methods for full-rank matrices cannot be applied.
When the coefficient matrix of the system is rank-deficient, 
the linear systems may be underdetermined, inconsistent, or both. 
In such cases, 
one generally seeks to compute the least squares solution 
that minimizes the residual of the problem,
and is further defined as the solution with smallest norm
in cases where the coefficient matrix has a nontrivial nullspace.

This work introduces several new techniques 
for solving least squares problems with coefficient matrices 
that are so large that 
they do not fit in main memory and 
must be stored in the disk drive.

All techniques rely on complete orthogonal factorizations
that guarantee that both conditions of a least squares solution
are met, regardless of the rank properties of the matrix.
That is, our new methods can solve linear systems
with both full-rank and rank-deficient coefficient matrices,
for both consistent and inconsistent problems.
Specifically, they rely on the recently proposed ``randUTV'' 
algorithm that is particularly effective in strongly 
communication-constrained environments.
Our work accelerates the speed of previous Out-Of-Core implementations
of the randUTV factorization
by introducing several performance improvements.
Moreover, new high-performance implementations for GPU were developed.

A detailed precision study of our new implementations has been performed,
which includes both our own matrices and
several large matrices from public repositories.
The precision of the new codes is competitive
with those of methods working in main memory.

A thorough performance study of the codes 
for solving the linear least squares problem has been performed.
The performances of the new codes, that operate on data stored on disk,
are competitive with current high-performance methods
that work only on data stored in main memory (thus limited in size).

The performances of our new codes are slower
than those of specific methods for full-rank matrices,
but only slightly so.
Our new methods can be employed both in CPU architectures
and GPU architectures,
delivering good performance and precision in both cases.

\section*{Acknowledgements}

G.~Quintana-Ort\'{\i}, M.~Chillar\'{o}n, and V.~Vidal were supported 
the TED2021-131091B-I00 research project, 
funded by MCIN/AEI/10.13039/501100011033 and by the 
``European Union NextGenerationEU/PRTR''.
M.~Chillar\'{o}n and V.~Vidal were also supported by ``Universitat Politècnica de València''.
P.G.~Martinsson acknowledges support from 
the Office of Naval Research (N00014-18-1-2354),
the National Science Foundation (DMS-2313434 and DMS-1952735),
the Department of Energy ASCR (DE-SC0022251), and
the Texas Advanced Computing Center.
The authors would like to thank 
Francisco D. Igual (Universidad Complutense de Madrid) 
for granting access to the \texttt{volta1} and the \texttt{mb} servers.


\nocite{golub}
\bibliography{main_bib}
\bibliographystyle{amsplain}

\end{document}